\RequirePackage{fix-cm}
\documentclass[twocolumn]{svjour3}          
\smartqed  

\usepackage{bm}
\usepackage{graphicx}
\usepackage{epstopdf}
\usepackage{multirow}
\usepackage{float}
\usepackage{threeparttable}
\usepackage{braket}
\usepackage{blkarray}
\usepackage{enumitem}
\usepackage{booktabs}
\usepackage{bigdelim}
\usepackage{ifthen}
\usepackage[utf8]{inputenc}
\usepackage{mathtools}
\usepackage{microtype}
\usepackage{standalone}
\usepackage{subcaption}
\usepackage{todonotes}
\usepackage{xargs}
\usepackage{xfrac}
\usepackage{xspace}
\usepackage{amsmath,amssymb}
\usepackage{algorithm}
\usepackage{caption}
\usepackage[noend]{algorithmic}
\usepackage{hyperref}
\usepackage[at]{easylist}
\usepackage{tikz}
\usepackage{xcolor}
\usepackage{color,xcolor,colortbl}
\usepackage{multirow}
\usepackage{breqn}
\usepackage{hhline}
\usepackage{tabularx}

\usetikzlibrary{shapes,arrows,shadows,positioning}
\newcommand{\overbar}[1]{\mkern 1.5mu\overline{\mkern-1.5mu#1\mkern-1.5mu}\mkern 1.5mu}

\NewDocumentCommand{\rot}{O{45} O{1em} m}{\makebox[#2][l]{\rotatebox{#1}{#3}}}

\definecolor{LightCyan}{rgb}{0.88,1,1}

\date{}

\begin{document}

\title{A Serial Multilevel Hypergraph Partitioning Algorithm
}

\author{Foad Lotfifar         \and
        Matthew Johnson 
}

\institute{School of Engineering and Computing Sciences, Durham University, South Rd, Durham, County Durham DH1 3LE, United Kingdom\\
              \email{foadfbf@gmail.com, matthew.johnson2@durham.ac.uk}
}

\maketitle

\begin{abstract}

The graph partitioning problem has many applications in scientific computing such as computer aided design, data mining, image compression and other applications with sparse-matrix vector multiplications as a kernel operation.  In many cases it is advantageous to use hypergraphs as they, compared to graphs, have a more general structure and can be used to model more complex relationships between groups of objects.   This motivates our focus on the  less-studied hypergraph partitioning problem.

In this paper, we propose a serial multi-level bipartitioning algorithm.  One important step in current heuristics for hypergraph partitioning is clustering during which similar vertices must be recognized.  This can be particularly difficult in irregular hypergraphs with high variation of vertex degree and hyperedge size;  heuristics that rely on local vertex clustering decisions often give poor partitioning quality.  A novel feature of the proposed algorithm is to use the techniques of rough set clustering to address this problem.  We show that our proposed algorithm gives on average between 18.8\% and 71.1\% better quality on these irregular hypergraphs by comparing it to state-of-the-art hypergraph partitioning algorithms on benchmarks taken from real applications.

\end{abstract}

\section{Introduction}\label{sec:intro}

This paper is concerned with hypergraph partitioning.  A hypergraph is a generalization of a graph in which edges (or hyperedges) can contain any number of vertices rather than just two.  Thus hypergraphs are able to represent more complex relationships and model less structured data~\cite{catayk1999,hendrickson1998graph,wang2014bilionnode} and thus can, for some applications in High Performance Computing, provide an improved model.
For this reason, hypergraph modelling and hypergraph partitioning have become widely used in the analysis of scientific applications~\cite{alp1996,catayk1999,curino2010schism,bey2014,hu2014,marquez2015,tian2009,zhou2006learning}. 

We propose a \textit{Feature Extraction Hypergraph Partitioning (FEHG)} algorithm for hypergraph partitioning which makes novel use of the technique of \textit{rough set clustering}.  We evaluate it in comparison with several state-of-the-art algorithms.

\subsection{The Hypergraph Partitioning Problem}\label{sec:hpart_problem}

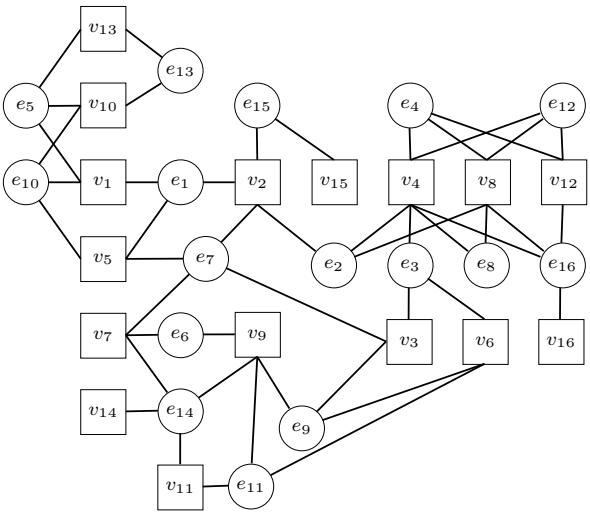
\begin{figure}[ht]
	\centering
	\tiny
	\captionsetup{font=small,labelfont=bf}
\resizebox {0.45\textwidth} {!} {
	\begin{tikzpicture}[outer sep=0.0cm,node distance=1.2cm,scale=\textwidth]
	\tikzstyle{vtx} = [draw=black,minimum size=0.7cm, rectangle,font=\footnotesize]
	\tikzstyle{nod} = [draw=black,minimum size=0.7cm, circle,font=\footnotesize]
	
	\node[vtx] 				(v1) 	{$v_{1}$};
	\node[vtx,above of=v1]	(v10)	{$v_{10}$};
	\node[vtx,above of=v10] 	(v13)	{$v_{13}$};
	\node[vtx,below of=v1]	(v5) 	{$v_{5}$};
	\node[vtx,below of=v5]	(v7) 	{$v_{7}$};
	\node[vtx,below of=v7]	(v14)	{$v_{14}$};

	\node[nod,left  =0.5cm of v10]	(e5)		{$e_{5}$};
	\node[nod,left  =0.5cm of v1]	(e10)	{$e_{10}$};
	\node[nod,below right=0.05cm and 0.6cm of v13](e13)	{$e_{13}$};
	\node[nod,right =0.5cm of v1]	(e1)		{$e_{1}$};

	\node[vtx,right =0.5cm of e1]	(v2)	{$v_{2}$};
	\node[vtx,right =0.5cm of v2]	(v15){$v_{15}$};
	\node[vtx,right of=v15]	(v4)		{$v_{4}$};
	\node[vtx,right of=v4]	(v8)		{$v_{8}$};
	\node[vtx,right of=v8]	(v12)	{$v_{12}$};

	\node[nod,below =0.6cm of v15]	(e2)		{$e_{2}$};
	\node[nod,below =0.6cm of v4]	(e3)		{$e_{3}$};
	\node[nod,below =0.6cm of v8]	(e8)		{$e_{8}$};
	\node[nod,below =0.6cm of v12]	(e16)	{$e_{16}$};
	\node[nod,above =0.5cm of v4]	(e4)		{$e_{4}$};
	\node[nod,above =0.5cm of v2]	(e15)	{$e_{15}$};
	\node[nod,above =0.5cm of v12]	(e12)	{$e_{12}$};
	\node[nod,right =0.5cm of v7]	(e6)		{$e_{6}$};
	\node[nod,right =0.5cm of v14]	(e14)	{$e_{14}$};
	
	\node[vtx,right =0.5cm of e6]	(v9)		{$v_{9}$};
	\node[vtx,below =0.5cm of e3]	(v3)		{$v_{3}$};
	\node[vtx,below =0.5cm of e8]	(v6)		{$v_{6}$};
	\node[vtx,below =0.5cm of e16]	(v16)	{$v_{16}$};
	\node[vtx,below =0.5cm of e14]	(v11)	{$v_{11}$};
	\node[nod,below left =0.6cm and 0.2cm of v2]	(e7)		{$e_{7}$};

	\node[nod,right =0.4cm of v11]	(e11)	{$e_{11}$};
	\node[nod,above right=0.4cm and 0.3cm of e11]	(e9)		{$e_{9}$};

	\draw (e5) -- (v13.west)			[draw=black,thick];
	\draw (e5) -- (v10.west)			[draw=black,thick];
	\draw (e5) -- (v1.west)			[draw=black,thick];
		
	\draw (e10) -- (v10.west)		[draw=black,thick];
	\draw (e10) -- (v1.west)			[draw=black,thick];
	\draw (e10) -- (v5.west)			[draw=black,thick];
	
	\draw (e13) -- (v13.east)		[draw=black,thick];
	\draw (e13) -- (v10.east)		[draw=black,thick];
	
	\draw (e1) -- (v1.east)			[draw=black,thick];
	\draw (e1) -- (v5.east)			[draw=black,thick];
	\draw (e1) -- (v2.west)			[draw=black,thick];
	
	\draw (e15) -- (v15.north)		[draw=black,thick];
	\draw (e15) -- (v2.north)		[draw=black,thick];

	\draw (e4) -- (v4.north)			[draw=black,thick];
	\draw (e4) -- (v8.north)			[draw=black,thick];
	\draw (e4) -- (v12.north)		[draw=black,thick];

	\draw (e12) -- (v4.north)		[draw=black,thick];
	\draw (e12) -- (v8.north)		[draw=black,thick];
	\draw (e12) -- (v12.north)		[draw=black,thick];

	\draw (e16) -- (v4.south)		[draw=black,thick];
	\draw (e16) -- (v8.south)		[draw=black,thick];
	\draw (e16) -- (v12.south)		[draw=black,thick];
	\draw (e16) -- (v16.north)		[draw=black,thick];

	\draw (e8) -- (v4.south)			[draw=black,thick];
	\draw (e8) -- (v8.south)			[draw=black,thick];

	\draw (e2) -- (v2.south)			[draw=black,thick];
	\draw (e2) -- (v4.south)			[draw=black,thick];
	\draw (e2) -- (v8.south)			[draw=black,thick];

	\draw (e3) -- (v4.south)			[draw=black,thick];
	\draw (e3) -- (v3.north)			[draw=black,thick];
	\draw (e3) -- (v6.north)			[draw=black,thick];

	\draw (e7) -- (v5.east)			[draw=black,thick];
	\draw (e7) -- (v2.south)			[draw=black,thick];
	\draw (e7) -- (v3.west)			[draw=black,thick];
	\draw (e7) -- (v7.east)			[draw=black,thick];

	\draw (e6) -- (v7.east)			[draw=black,thick];
	\draw (e6) -- (v9.west)			[draw=black,thick];

	\draw (e14) -- (v7.east)		[draw=black,thick];
	\draw (e14) -- (v14.east)		[draw=black,thick];
	\draw (e14) -- (v9.south)		[draw=black,thick];
	\draw (e14) -- (v11.north)		[draw=black,thick];

	\draw (e11) -- (v9.south)		[draw=black,thick];
	\draw (e11) -- (v11.east)		[draw=black,thick];
	\draw (e11) -- (v6.south)		[draw=black,thick];

	\draw (e9) -- (v9.south)			[draw=black,thick];
	\draw (e9) -- (v3.west)			[draw=black,thick];
	\draw (e9) -- (v6.south)			[draw=black,thick];

	\end{tikzpicture}
}
\caption{A sample hypergraph $H$ with $16$ vertices and 16 hyperedges. Vertices and hyperedges are represented as square and circular nodes, respectively. The weights of the vertices and hyperedges are assumed to be 1. }\label{fig:fehg_sample}
\end{figure}

We first define the problem formally.  A hypergraph $H=(V,E)$ is a finite set of vertices $V$ and a finite set $E \subseteq 2^V$ of hyperedges.  Each hyperedge $e \in E$ can contain any number of vertices (constrast with a graph where every edge contains two vertices). Let $e \in E$ and $v \in V$ be a hyperedge and a vertex of $H$, respectively. Then hyperedge $e$ is said to be \textit{incident} to $v$ or to \textit{contain} $v$, if $v \in e$.  This is denoted $e \triangleright v$. The pair $\langle e,v \rangle$ is further called a \textit{pin} of $H$. The degree of $v$ is the number of hyperedges incident to $v$ and is denoted $d(v)$. The size or cardinality of a hyperedge $e$ is the number of vertices it contains and is denoted $|e|$. An example of a hypergraph with 16 vertices and 16 hyperedges is given in Fig.~\ref{fig:fehg_sample}.

Let $\omega \colon V \mapsto \mathbb{N}$ and $\gamma \colon E \mapsto \mathbb{N}$ be functions defined on the vertices and hyperedges of $H$.  For $v \in V$ and $e \in E$, we call $\omega(v)$ and $\gamma(e)$ the \textit{weights} of $v$ and~$e$.   For a non-negative integer $k$, a \textit{k-way partitioning} of $H$ is a collection of non-empty disjoint sets $\Pi = \left\lbrace \pi_1, \pi_2, \cdots, \pi_k \right\rbrace$  such that $\bigcup_{i=1}^{k} \pi_i=V$.   If $k=2$, we call~$\Pi$ a bipartition.  We say that vertex $v \in V$ is \emph{assigned} to a part $\pi$ if $v \in \pi$. Furthermore, the weight of a part $\pi \in \Pi$ is the sum of the weights of the vertices assigned to that part. A hyperedge $e \in E$ is said to be \textit{connected} to (or \textit{spans} on) the part $\pi$ if $e \cap \pi \neq \emptyset$. The \textit{connectivity degree} of a hyperedge $e$ is the number of parts connected to $e$ and is denoted $\lambda_e(H,\Pi)$. A hyperedge is said to be \textit{cut} if its connectivity degree is more than 1.

We define the \textit{partitioning cost} of a $k$-way partitioning $\Pi$ of a hypergraph $H$:

\begin{equation} \label{eq:costfunction}
\operatorname{cost}(H,\Pi)=\sum\nolimits_{e \in E} (\gamma(e) \cdot  (\lambda_{e} (H,\Pi)-1))
\end{equation}

The \emph{imbalance tolerance} is a real number $\epsilon \in (0,1)$.   Then a $k$-way partitioning $\Pi$ of a hypergraph $H$ satisfies the \textit{balance constraint} if

\begin{equation}\label{eq:imbalance}
\overbar{W} \cdot (1 - \epsilon) \leqslant \omega(\pi) \leqslant \overbar{W} \cdot (1 + \epsilon), \; \forall \pi \in \Pi
\end{equation}
where $\overbar{W} =\sum\nolimits_{v \in V} \omega(v)/k$ is the average weight of a part.

The \textit{Hypergraph Partitioning Problem} is to find a $k$-way partitioning with minimum cost subject to the satisfaction of of the balance constraint.

\subsection{Heuristics for Hypergraph Partitioning}\label{sec:relatedwork}

The Hypergraph Partitioning Problem is known to be NP-Hard \cite{garey1979computers}, but many good heuristic algorithms have been proposed to solve the problem \cite{ccatalyurek2011patoh,devetal2006,fm1982,karypis1999kway} and we now briefly give an overview of the methods that have been developed.

As graph partitioning is a well-studied problem and good heuristic algorithms are known~\cite{fjall1998graphsurvey} one might expect to solve the hypergraph partitioning problem using graph partitioning. This requires a transformation of the hypergraph into a graph that preserves its structure. There is no known algorithm for this purpose \cite{ihler1993}.

\textit{Move-based} or \textit{flat} algorithms start with an initialisation step in which vertices are assigned at random to the parts. Then the algorithm builds a new problem solution based on the neighbourhood structure of the hypergraph~\cite{fm1982,san1989kfm}. The main weakness of these algorithms is that while the solution found is locally optimal, whether or not it is also globally optimal seems to depend on the density of the hypergraph and the initial random distribution~\cite{goldberg1983}. Saab and Rao \cite{saab1992rao} show that the performance of one such algorithm (the KL algorithm) improves as the graph density increases.

\textit{Recursive bipartitioning} algorithms generate a bipartitioning of the original hypergraph and are then recursively applied independently to both parts until a $k$-way partitioning is obtained \cite{fm1982,ccatalyurek2011patoh,devetal2006}. {Direct $k$-way} algorithms calculate a $k$-way partitioning by working directly on the hypergraph \cite{san1989kfm}. There is no consensus on which paradigm gives better partitioning. Karypis reports some advantages of direct {$k$-way} partitioning over recursive algorithms \cite{karytech2002}. Cong et al.  report that direct algorithms are more likely to get stuck in local minima~\cite{cong1998}. Aykanat et al.  report that direct algorithms give better results for large $k$~\cite{aykanat2008fixed}. In fact, recursive algorithms are widely used in practice and are seen to achieve good performance and quality. 

We mention in passing that other algorithms consider the hypergraph partitioning problem with further restrictions. Multi-constraint algorithms, are used when more than one partitioning objective is defined by assigning a weight vector to the vertices \cite{aykanat2008fixed}.   In the context of VLSI circuit partitioning, there are terminals, such as I/O chips, that are fixed and cannot be moved. Therefore when partitioning these circuits some vertices are fixed and must be assigned to particular parts. Problems of this type are easier to solve and faster running times can be achieved~\cite{alpert2000fixed,ccatalyurek2011patoh,aykanat2008fixed}. 

In this paper, we introduce a \textit{multi-level} algorithm so let us describe this general approach.  These algorithms have three distinct phases: \textit{coarsening}, \textit{initial partitioning} and \textit{uncoarsening}. During coarsening vertices are merged to obtain hypergraphs with progressively smaller vertex sets. After the coarsening stage, the partitioning problem is solved on the resulting smaller hypergraph. Then in the uncoarsening stage, the coarsening stage is reversed and the solution obtained on the small hypergraph is used to provide a solution on the input hypergraph. The coarsening phase is sometimes also known as the \textit{refinement phase}.

During coarsening the focus is on finding clusters of vertices to merge to form vertices in the coarser hypergraph.  One wishes to merge vertices that are, in some sense, alike so a metric of similarity is required; examples are Euclidean distance, Hyperedge Coarsening, First Choice (FC), or the similarity metrics proposed by Alpert et al. \cite{alpert1998multilevel}, and Catalyurek and Aykanat \cite{catayk1999}. 

A problem encountered with high dimensional data sets is that the similarity between objects becomes very non-uniform and defining a similarity measure and finding similar vertices to put into one cluster is very difficult.  This situation is more probable when the mean and standard deviation of the vertex degrees increases (see, for example, the analysis of {Euclidean distance} as a similarity measures by Ert{\"o}z et al. \cite{ertozetal2003}).  Although other measures, such as \textit{cosine measure} and \textit{jaccard distance}, address the issue and resolve the problem to some extent, they also have limitations. For example, Steinbach et al. \cite{steinbach2000} have evaluated these measures and found that they fail to capture similarity between text documents in document clustering techniques (used in areas such as text mining and information retrieval). The problem is that cosine and jaccard distances emphasise the importance of shared attributes for measuring similarity and they ignore attributes that are not common to pairs of vertices. Consequently, other algorithms use other clustering techniques to resolve the problem such as \textit{shared nearest neighbour (SNN)} methods \cite{ertoz2002new} and global vertex clustering information. 

Decisions for vertex clustering are made locally and global decisions are avoided due to their high cost and complexity though they give better results \cite{ale2006}. All proposed heuristics reduce the search domain and try to find vertices to be matched using some degree of randomness \cite{devetal2006}. This degrades the quality of the partitioning by increasing the possibility of getting stuck in a local minimum and they are highly dependent on the order the vertices are selected for matching. A better trade-off is needed between the low cost of local decisions and the high quality of global ones.

Furthermore, there is a degree of redundancy in modelling scientific applications with hypergraphs. Removing this redundancy can help in some optimisations such as clustering decisions, storage overhead, and processing requirements. An example is proposed by Heinz and Chandra \cite{bey2014} in which the hypergraph is transformed into a Hierarchy DAG (HDAG) representation and reducing the memory requirement for storing hypergraphs. Similarly, one can identify redundancies in the coarsening phase and make better vertex clustering decisions and achieve better partitioning on the hypergraph. 

The \textit{FEHG} algorithm proposed in this paper makes novel use of the technique of rough set clustering to categorise the vertices of a hypergraph in the coarsening phase. \textit{FEHG} treats hyperedges as features of the hypergraph and tries to discard unimportant features to make better clustering decisions. It also focuses on the trade-off to be made between local vertex matching decisions (which have low cost in terms of the space required and time taken) and global decisions (which can be of better quality but have greater costs). The emphasis of our algorithm is on the coarsening phase: good vertex clustering decisions here can lead to high quality partitioning~\cite{karytech2002}. 

\section{Preliminaries}\label{sec:preliminaries}

\subsection{Rough Set Clustering}\label{sec:rough_sets}

Rough set clustering is a mathematical approach to uncertainty and vagueness in data analysis. The idea was first introduced by Pawlak in 1991 \cite{pawlak1991}. The approach is different from statistical approaches, where the probability distribution of the data is needed, and fuzzy logic, where a degree of membership is required for an object to be a member of a set or cluster. The approach is based on the idea that every object in a universe is tied with some knowledge or attributes such that objects which are tied to the same attributes are \textit{indiscernible} and can be put together in one category \cite{pawlak2005rough}. 

The data to be classified are called \textit{objects} and they are described in an \textit{information system}:

\begin{definition} [Information System]\label{def:infosys}
An \textit{information system} is a system represented as $\mathfrak{I}=(\mathbb{U},\mathbf{A},\mathbf{V},\mathcal{F} )$ where:
	\begin{easylist}[itemize]
	@ \emph{$\mathbb{U}$} is a non-empty finite set of objects or the universe.
	@ \emph{$\mathbf{A}$} is a non-empty finite set of attributes with size $\vert \mathbf{A} \vert = t$.
	@ \emph{$\mathbf{V} = (\mathbf{V}_1, \mathbf{V}_2, \ldots, \mathbf{V}_t)$} where, for each $a \in \mathbf{A}$, $\mathbf{V}_a$ is the set of values for $a$.
	@ \emph{$\mathcal{F}: \mathbb{U} \times \mathbf{A} \mapsto \mathbf{V}$} is a mapping function such that, for $u \in \mathbb{U}, a \in \mathbf{A}$, $\mathcal{F}(u,a) \in \mathbf{V}_a$.  
	\end{easylist}
\end{definition}


For any subset of attributes $\mathbf{B} = \left\lbrace b_1,b_2,\cdots,b_j \right\rbrace \subseteq \mathbf{A}$, the \textit{B-Indiscernibility relation} is defined as follow:

\begin{equation}\label{ch2:eq:indis}
	\text{IND}(\mathbf{B})=\left\lbrace (u,v) \in \mathbb{U}^2 \mid \forall b \in 	\mathbf{B}, \,\mathcal{F}(u,b)=\mathcal{F}(v,b) \right\rbrace
\end{equation}

When $(u,v) \in \text{IND}(\mathbf{B})$, it is said that $u$ and $v$ are indiscernible under $B$. The equivalence class of $u$ with respect to $B$ is represented as $[u]_{\mathbf{B}}$ and includes all objects which are indiscernible with $u$. The equivalence relation provides a partitioning of the universe $\mathbb{U}$ in which each object belongs only one part and it is denoted $\mathbb{U}/\text{IND}(\mathbf{B})$ or simply $\mathbb{U} / \mathbf{B}$.

The set of attributes can contain some redundancy. Removing this redundancy could lead us to a better clustering decisions and data categorisation while still preserving the indiscernibility relation amongst the objects \cite{wroblewski1998,ziarko1995}. The remained attributes after removing the redundancy is called the \textit{reduct} set \cite{roughreview2009}.  More precisely, if $\mathbf{B} \subseteq \mathbf{A}$, then $\mathbf{B}$ is a \textbf{reduct} of $\mathbf{A}$ if $\mbox{IND}(\mathbf{B}) = \mbox{IND}(\mathbf{A})$ and $\mathbf{B}$ is \emph{minimal} (that is, no attribute can be removed from $\mathbf{B}$ without changing the indiscernibility relation).  But the reduct is not unique and it is known that 
finding a reduct of an information system is an NP-hard problem \cite{skowron1992}. This is one of the computational bottlenecks of rough set 
clustering. A number of heuristic algorithms have been proposed for problems where the number of attributes is small. Examples have been given by Wroblewski using genetic algorithms~\cite{wroblewski1995,wroblewski1998}, and by Ziarko and Shan who use decision tables based on Boolean algebra \cite{ziarko1995}. These methods are not applicable to hypergraphs which usually represent applications with high dimensionality and where the operations have to be repeated several time during partitioning. We propose a relaxed feature reduction method for hypergraphs by defining the Hyperedge Connectivity Graph (HCG).

\subsection{The Hyperedge Connectivity Graph}\label{sec:hcg}

The \textit{Hyperedge Connectivity Graph (HCG)} of a hypergraph is the main tool used used in our algorithm for removing superfluous and redundant information during vertex clustering in the coarsening phase.  The similarity between two hyperedges $e_i$ and $e_j$ is denoted $sim(e_i \cdot e_j)$ (we will discuss later possible similarity measures). 

\begin{definition} [Hyperedge Connectivity Graph] \label{def:hcg}
For a given \textit{similarity threshold} $s \in (0,1)$, the Hyperedge Connectivity Graph (HCG) of a hypergraph $H=(V,E)$ is a graph $\mathcal{G}^s=(\mathcal{V},\mathcal{E})$ where $\mathcal{V}=E$ and two hyperedges $e_i,e_j \in \mathcal{V}$ are adjacent in $\mathcal{G}^s$ if $sim(e_i,e_j) \geqslant s$.
\end{definition}

The definition is similar to that of \textit{intersection graphs} \cite{erdos1966representation} (graphs that represent the intersections of a family of sets) or  \textit{line graphs} of  hypergraphs (graphs whose vertex set is the set of the hyperedges of the hypergraph and two hyperedges are adjacent when their intersection is non-empty). The difference is the presence of the similarity measure that reduces the number of edges in the HCG. Different similarity measures, such as \textit{Jaccard Index} or \textit{Cosine Measure}, can be used for measuring the similarity. As the hyperedges of the hypergraph are weighted, similarity between two hyperedges is scaled according to the weight of hyperedges: for $e_i,e_j \in E$ the scaling factor is $\frac{\gamma(e_i) + \gamma(e_j)} {2 \times \max_{e \in E}\left( \gamma(e) \right) }$. One of the characteristics of the HCG is that it assigns hyperedges to non-overlapping clusters (that is, the connected components of the HCG).

\section{The Serial Partitioning Algorithm}\label{sec:fehg_alg}

The proposed Feature Extraction Hypergraph Partitioning (\textit{FEHG}) algorithm is a multi-level recursive serial bipartitioning algorithm composed of three distinct phases: coarsening, initial partitioning, and uncoarsening. The emphasis of \textit{FEHG }is on the coarsening phase as it is the most important phase of the multi-level paradigm~\cite{karytech2002}. 

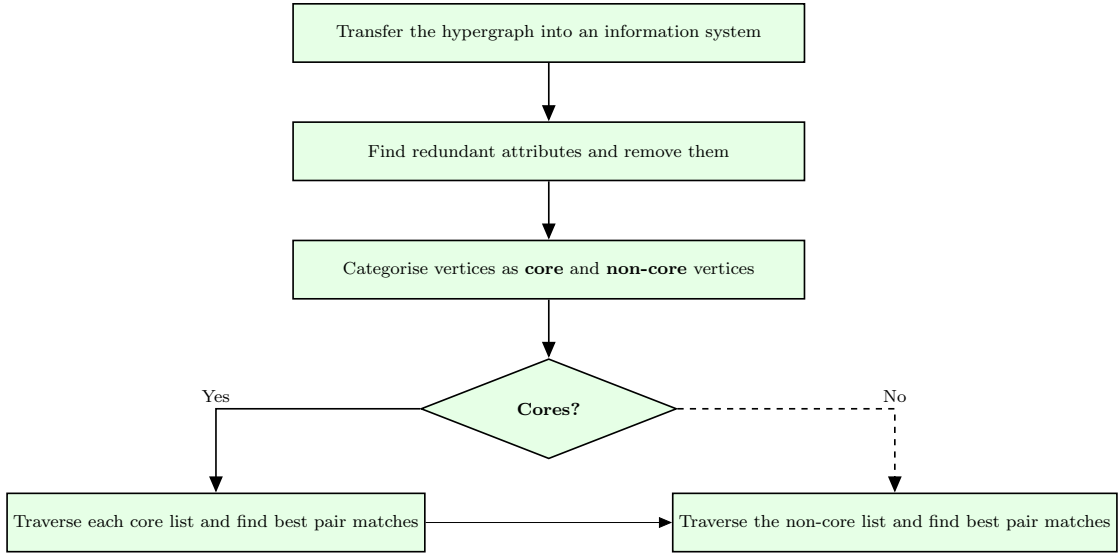
\begin{figure*}[ht]
	\centering
	\captionsetup{labelfont=bf}
    \resizebox {0.85\textwidth} {!} {
    	\begin{tikzpicture}
	[align=center]
	
	\node (transfer) 	[draw, thick, fill=green!10, shape=rectangle, minimum width=0.5\textwidth, minimum height=1cm] {Transfer the hypergraph into an information system};

	\node (redattr) 	[draw, thick, fill=green!10, shape=rectangle, minimum width=0.5\textwidth, minimum height=1cm, below= of transfer] {Find redundant attributes and remove them};

	\node (cores) 	[draw, thick, fill=green!10, shape=rectangle, minimum width=0.5\textwidth, minimum height=1cm, below= of redattr] {Categorise vertices as \textbf{core} and \textbf{non-core} vertices};
		
	\node (decide) 	[draw, thick, fill=green!10, shape=diamond, minimum width=0.25\textwidth, minimum height=1cm, below= of cores] {\textbf{Cores?}};

	\node (tracores) 	[draw, thick, fill=green!10, shape=rectangle, minimum width=0.25\textwidth, minimum height=1cm, below left= of decide] {Traverse each core list and find best pair matches};

	\node (tranoncores) 	[draw, thick, fill=green!10, shape=rectangle, minimum width=0.25\textwidth, minimum height=1cm, below right= of decide] {Traverse the non-core list and find best pair matches};

	\draw [-triangle 45] (transfer.south) 	to (redattr.north)		[draw,thick];
	\draw [-triangle 45](redattr.south) 	to (cores.north)			[draw,thick];
	\draw [-triangle 45](cores.south) 		to (decide.north)		[draw,thick];
	\draw [-triangle 45](decide.west) 	-| (tracores.north)[thick] node [pos=0.5,above]{Yes};
	\draw [dashed,-triangle 45](decide.east) 	-| (tranoncores.north)[thick] node [pos=0.5,above]{No};
	\draw [-triangle 45](tracores.east) 	to (tranoncores.west) node [pos=0.5,above]{};
	
	\end{tikzpicture}
	}
	\caption{The coarsening phase at a glance. The dashed arrow means that non-core vertex list is processed after all cores have been processed.}\label{fig:coarse_glance}
\end{figure*}

We provide a general outline of the algorithm before going into further detail. In the coarsening phase, we transfer the hypergraph into an information system and we use rough set clustering decisions to match pairs of vertices. This is done in several steps. First, the reduct set is found to reduce the size of the system and remove superfluous information. Second, the vertices of the hypergraph are categorised using their indispensability relations. These categories are denoted as \textit{core} and \textit{non-core} where, in some sense, vertices that belong to the same core are good candidates to be matched together and the non-core vertices are what is left over. The cores are traversed before the non-cores for finding pair-vertex matches. The whole coarsening procedure is depicted in Fig.~\ref{fig:coarse_glance}.

\subsection{The Coarsening}\label{sec:coarsening}

The first step of the coarsening stage is to transfer the hypergraph $H=(V,E)$ into an information system. The information system representing the hypergraph is $\mathcal{I}_{H}=(V,E,\mathbf{V},\mathcal{F})$; that is, the objects are the vertices$V$ and the attributes are the hyperedges $E$. We define the set of values as each being in $[0,1]$ and the mapping function is defined as:

\begin{equation}\label{eq:hgraph_info_mapping}
\mathcal{F}(v,e) = \frac{f(e)}{\sum\nolimits_{\forall e' \triangleright v} \gamma(e')}, 
\end{equation}%
where $f(e)=\gamma(e)$ if $e \triangleright v$ and is otherwise 0.

The transformation of the hypergraph given in Fig.~\ref{fig:fehg_sample} into an information system is given in Table~\ref{tab:hgraph_into_infosys}. 

\begin{table*}[ht]
	\centering
	\captionsetup{labelfont=bf}
	\caption{The transformation of the hypergraph $H$ depicted in Fig.~\ref{fig:fehg_sample} into an information system. The values are rounded to 2 decimal places. } \label{tab:hgraph_into_infosys}
	\resizebox{0.9\textwidth}{!}{%
	\begin{threeparttable}
		\begin{tabular}{ccccccccccccccccc}	
		\hline
		\rowcolor{LightCyan}
					 		& $e_1$ 		& $e_2$ 		& $e_3$ 		& $e_4$
					 		& $e_5$ 		& $e_6$ 		& $e_7$ 		& $e_8$ 
					 		& $e_9$ 		& $e_{10}$	& $e_{11}$	& $e_{12}$
					 		& $e_{13}$	& $e_{14}$	& $e_{15}$	& $e_{16}$ \\
		\hline
		$\mathbf{v_1}$	 	& 	$0.33$	&	$0  $	&	$0  $	& 	$0  $
							& 	$0.33$	&	$0  $	&	$0  $	& 	$0  $
							& 	$0  $	&	$0.33$	&	$0  $	& 	$0  $
							& 	$0  $	&	$0  $	&	$0  $	& 	$0  $	\\
							
		$\mathbf{v_2}$ 		& 	$0.25$	&	$0.25$	&	$0  $	& 	$0  $
							& 	$0  $	&	$0  $	&	$0.25$	& 	$0  $
							& 	$0  $	&	$0  $	&	$0  $	& 	$0  $
							& 	$0  $	&	$0  $	&	$0.25$	& 	$0  $	\\
							
		$\mathbf{v_3}$ 		& 	$0  $	&	$0  $	&	$0.33$	& 	$0  $
							& 	$0  $	&	$0  $	&	$0.33$	& 	$0  $
							& 	$0.33$	&	$0  $	&	$0  $	& 	$0  $
							& 	$0  $	&	$0  $	&	$0  $	& 	$0  $	\\
							
		$\mathbf{v_4}$ 		& 	$0  $	&	$0.16$	&	$0.16$	& 	$0.16$
							& 	$0  $	&	$0  $	&	$0  $	& 	$0.16$
							& 	$0  $	&	$0  $	&	$0  $	& 	$0.16$
							& 	$0  $	&	$0  $	&	$0  $	& 	$0.16$	\\
							
		$\mathbf{v_5}$ 		& 	$0.33$	&	$0  $	&	$0  $	& 	$0  $
							& 	$0  $	&	$0  $	&	$0.33$	& 	$0  $
							& 	$0  $	&	$0.33$	&	$0  $	& 	$0  $
							& 	$0  $	&	$0  $	&	$0  $	& 	$0  $	\\
							
		$\mathbf{v_6}$ 		& 	$0  $	&	$0  $	&	$0.33$	& 	$0  $
							& 	$0  $	&	$0  $	&	$0  $	& 	$0  $
							& 	$0.33$	&	$0  $	&	$0.33$	& 	$0  $
							& 	$0  $	&	$0  $	&	$0  $	& 	$0  $	\\
							
		$\mathbf{v_7}$ 		& 	$0  $	&	$0  $	&	$0  $	& 	$0  $
							& 	$0  $	&	$0.33$	&	$0.33$	& 	$0  $
							& 	$0  $	&	$0  $	&	$0  $	& 	$0  $
							& 	$0  $	&	$0.33$	&	$0  $	& 	$0  $	\\
							
		$\mathbf{v_8}$ 		& 	$0  $	&	$0.2$	&	$0  $	& 	$0.2$
							& 	$0  $	&	$0  $	&	$0  $	& 	$0.2$
							& 	$0  $	&	$0  $	&	$0  $	& 	$0.2$
							& 	$0  $	&	$0  $	&	$0  $	& 	$0.2$	\\
							
		$\mathbf{v_9}$ 		& 	$0  $	&	$0  $	&	$0  $	& 	$0  $
							& 	$0  $	&	$0.25$	&	$0  $	& 	$0  $
							& 	$0.25$	&	$0  $	&	$0.25$	& 	$0  $
							& 	$0  $	&	$0.25$	&	$0  $	& 	$0  $	\\
							
		$\mathbf{v_{10}}$ 	& 	$0  $	&	$0  $	&	$0  $	& 	$0  $
							& 	$0.33$	&	$0  $	&	$0  $	& 	$0  $
							& 	$0  $	&	$0.33$	&	$0  $	& 	$0  $
							& 	$0.33$	&	$0  $	&	$0  $	& 	$0  $	\\
							
		$\mathbf{v_{11}}$ 	& 	$0  $	&	$0  $	&	$0  $	& 	$0  $
							& 	$0  $	&	$0  $	&	$0  $	& 	$0  $
							& 	$0  $	&	$0  $	&	$0.5$	& 	$0  $
							& 	$0  $	&	$0.5$	&	$0  $	& 	$0  $	\\
							
		$\mathbf{v_{12}}$ 	& 	$0  $	&	$0  $	&	$0  $	& 	$0.33$
							& 	$0  $	&	$0  $	&	$0  $	& 	$0  $
							& 	$0  $	&	$0  $	&	$0  $	& 	$0.33$
							& 	$0  $	&	$0  $	&	$0  $	& 	$0.33$	\\
							
		$\mathbf{v_{13}}$ 	& 	$0  $	&	$0  $	&	$0  $	& 	$0  $
							& 	$0.5$	&	$0  $	&	$0  $	& 	$0  $
							& 	$0  $	&	$0  $	&	$0  $	& 	$0  $
							& 	$0.5$	&	$0  $	&	$0  $	& 	$0  $	\\
							
		$\mathbf{v_{14}}$ 	& 	$0  $	&	$0  $	&	$0  $	& 	$0  $
							& 	$0  $	&	$0  $	&	$0  $	& 	$0  $
							& 	$0  $	&	$0  $	&	$0  $	& 	$0  $
							& 	$0  $	&	$1.0$	&	$0  $	& 	$0  $	\\
							
		$\mathbf{v_{15}}$ 	& 	$0  $	&	$0  $	&	$0  $	& 	$0  $
							& 	$0  $	&	$0  $	&	$0  $	& 	$0  $
							& 	$0  $	&	$0  $	&	$0  $	& 	$0  $
							& 	$0  $	&	$0  $	&	$1.0$	& 	$0  $	\\
							
		$\mathbf{v_{16}}$ 	& 	$0  $	&	$0  $	&	$0  $	& 	$0.25$
							& 	$0  $	&	$0  $	&	$0  $	& 	$0.25$
							& 	$0  $	&	$0  $	&	$0  $	& 	$0.25$
							& 	$0  $	&	$0  $	&	$0  $	& 	$0.25$	\\
		
		\bottomrule
		\end{tabular}
	\end{threeparttable}
	} 
\end{table*}

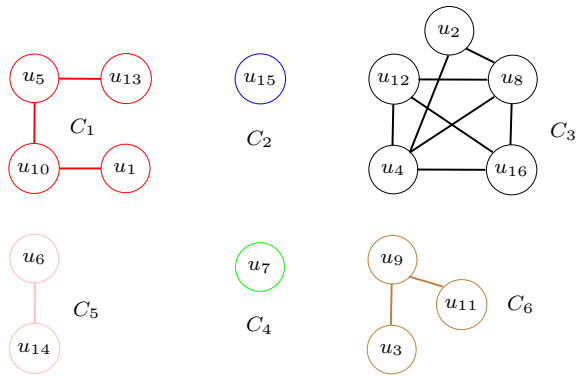
\begin{figure}[ht]
	\centering
	\tiny
	\captionsetup{font=small,labelfont=bf}
\resizebox {0.45\textwidth} {!} {
	\begin{tikzpicture}[outer sep=0.0cm,node distance=1.3cm,scale=\textwidth]
	\tikzstyle{ecn} = [minimum size=0.7cm, circle,font=\footnotesize]
	\tikzstyle{noframe} = [minimum size=0.8cm, rectangle,font=\footnotesize]
	
	\node[ecn,draw=red]						(u1)		{$u_{1}$};
	\node[ecn,draw=red,left =0.6cm of u1]	(u10)	{$u_{10}$};
	\node[ecn,draw=red,above of= u10]		(u5)		{$u_{5}$};
	\node[ecn,draw=red,right =0.6cm of u5]	(u13)	{$u_{13}$};
	\node[noframe,below right=0.05cm and 0.05cm of u5] (c1)	{$C_{1}$};
	\draw (u1.west)  	-- (u10.east)		[draw=red,thick];
	\draw (u10.north)  	-- (u5.south)		[draw=red,thick];
	\draw (u13.west)  	-- (u5.east)			[draw=red,thick];
	
	\node[ecn,draw=blue,right = 1.2cm of u13](u15)	{$u_{15}$};
	\node[noframe,below = 0.1cm of u15]		(c2)		{$C_{2}$};

	\node[ecn,draw=black,right =1.2cm of u15](u12)	{$u_{12}$};
	\node[ecn,draw=black,right =1cm of u12](u8)	{$u_{8}$};
	\node[ecn,draw=black,below of= u8]		(u16)	{$u_{16}$};
	\node[ecn,draw=black,below of= u12]		(u4)		{$u_{4}$};
	\node[ecn,draw=black,above right=0.2cm and 0.3cm of u12](u2)	{$u_{2}$};
	\node[noframe,below right=0.1cm and 0.1cm of u8]	(c3)	{$C_{3}$};
	\draw (u2.south east) 	-- (u8.north west)			[draw=black,thick];
	\draw (u2.south) 	-- (u4.north east)	[draw=black,thick];
	\draw (u8.west) 		--  (u12.east)		[draw=black,thick];
	\draw (u8.south west) 		--  (u4.north east)		[draw=black,thick];
	\draw (u8.south) 	--  (u16.north)		[draw=black,thick];
	\draw (u12.south) 	--  (u4.north)		[draw=black,thick];
	\draw (u16.west) 	-- (u4.east)			[draw=black,thick];
	\draw (u12.south east) 	-- (u16.north west)		[draw=black,thick];

	\node[ecn,draw=pink,below of= u10]		(u6)		{$u_{6}$};
	\node[ecn,draw=pink,below of= u6]		(u14)	{$u_{14}$};
	\node[noframe,below right =0.1cm and 0.1cm of u6]	(c5)	{$C_{5}$};
	\draw (u6.south) 	-- (u14.north)		[draw=pink,thick];

	\node[ecn,draw=green,below =2cm of u15]	(u7)		{$u_{7}$};
	\node[noframe,below = 0.1cm of u7]		(c4)		{$C_{4}$};

	\node[ecn,draw=brown,below of= u4]		(u9)		{$u_{9}$};
	\node[ecn,draw=brown,below of= u9]		(u3)		{$u_{3}$};
	\node[ecn,draw=brown,below right = 0.15cm and 0.5cm of u9](u11)	{$u_{11}$};
	\node[noframe,right =0.1cm of u11]		(c6)		{$C_{6}$};
	\draw (u9.south) 	-- (u3.north)		[draw=brown,thick];
	\draw (u9.south east) 	-- (u11.north west)		[draw=brown,thick];

	\end{tikzpicture}
}
\caption{An example of Hyperedge Connectivity Graph (HCG) of the hypergraph $H$ depicted in Fig.~\ref{fig:fehg_sample} and the similarity threshold $s=0.5$.}\label{fig:hcg_sample}
\end{figure}

Calculating the HCG: Initially, hyperedges are not assigned to any cluster.  Then, until all hyperedges are assigned,  a hyperedge $e$ is selected, a new cluster number is assigned to it and the cluster is developed around $e$ by processing its adjacent hyperedges using Definition~\ref{def:hcg}. At the end of HCG calculation, each hyperedge is assigned to exactly one cluster. We refer to the cluster set as \textit{edge partitions} and denote it $E^{\mathrm{R}}$. The size and weight of each $e_{\mathrm{R}} \in E^{\mathrm{R}}$ is the number of hyperedges it contains and the sum of their weights, respectively. An example of the HCG for the sample hypergraph in Fig.~\ref{fig:fehg_sample} and similarity threshold $s=0.5$ is depicted in Fig.~\ref{fig:hcg_sample}.

Hyperedges belonging to the same edge partition are considered to be  similar. For each edge partition, we choose a representative and all other hyperedges in the same edge partition are removed from the information system $\mathcal{I}_{H}$ and replaced by their representative. After this reduction, the new information system is built and represented as $\mathcal{I}^{\mathrm{R}}_{H} = \left( V,E^{\mathrm{R}},\mathbf{V}^{\mathrm{R}},\mathcal{F}^{\mathrm{R}} \right)$ in which the attribute set is replaced by $E^{\mathrm{R}}$. In the new information system, the set of values is $\mathbf{V}^{\mathrm{R}}_{e_{\mathrm{R}}} \subseteq \mathbb{N}$. Furthermore, the mapping function for $\forall e_{\mathrm{R}} \in E^{\mathrm{R}}$ is redefined as follows:

\begin{equation}
\label{eq:epart_map}
\mathcal{F}^{\mathrm{R}}(v,e_{\mathrm{R}})=\left\vert {\left\lbrace e \triangleright v \, \wedge \, e \in e_{\mathrm{R}} |  e \in E \ \right\rbrace} \right\vert.
\end{equation}
We can further reduce the set of attributes by picking the most important ones. For this purpose, we define a \textit{clustering threshold} $c \in [0,1]$ and the mapping function is changed according to Eq.(\ref{eq:finalmap}) below to construct the final information system $\mathcal{I}^{\mathrm{f}}$. 

\begin{equation}\label{eq:finalmap}
\mathcal{F}^f(v,e_{\mathrm{R}}) = 
\begin{cases}
 1, & \mbox{if} \, \frac{\mathcal{F}^{\mathrm{R}}(v,e_{\mathrm{R}})}{\mid \left\lbrace e \triangleright v, \forall e \in E\right\rbrace \mid} \geqslant c\\
 0, & \mbox{otherwise}.
\end{cases}
\end{equation}

An example of the reduced information system and the final table using the clustering threshold $c=0.5$ for the hypergraph of Fig.~\ref{fig:fehg_sample} is depicted in Table~\ref{table:reduct_infosys2}. The final table is very sparse compared to the original table. At this point, we use rough set clustering. For every vertex, we calculate its equivalence class. Then, a partitioning $\mathbb{U}/\text{IND}(E^{\mathrm{R}})$ on the vertex set is obtained using the equivalence relations. We refer to parts in $\mathbb{U}/\text{IND}(E^{\mathrm{R}})$ as \textit{cores} such that each vertex belongs to a unique core. For some of the vertices in the hypergraph, the mapping function gives zero output for all attributes that is $\mathcal{F}^{\mathrm{f}}(v,e_{\mathrm{R}})=0, \, \forall e_{\mathrm{R}} \in E^{\mathrm{R}}$. These vertices are assigned to a list denoted as the \textit{non-core} vertex list. 

\begin{table}[t]
	\centering
	\caption{The reduced information system that is built based on the HCG in Fig.~\ref{fig:hcg_sample}.}\label{table:reduct_infosys1}

	\resizebox{0.8\linewidth}{!}{%
	\begin{threeparttable}
		\begin{tabular}{|ccccccc|}
		\hline
		\rowcolor{LightCyan}
					 		& $C_1$ 		& $C_2$ 		& $C_3$ 		& $C_4$
					 		& $C_5$ 		& $C_6$	\\
		\hline
		$\mathbf{v_1}$	 	& $3$	& $0$	& $0$	& $0$	& $0$	& $0$	\\
							
		$\mathbf{v_2}$ 		& $1$	& $1$	& $1$	& $1$	& $0$	& $0$	\\
							
		$\mathbf{v_3}$ 		& $0$	& $0$	& $1$	& $1$	& $0$	& $2$	\\
							
		$\mathbf{v_4}$ 		& $1$	& $0$	& $3$	& $0$	& $0$	& $0$	\\
							
		$\mathbf{v_5}$ 		& $2$	& $0$	& $0$	& $1$	& $0$	& $0$	\\
							
		$\mathbf{v_6}$ 		& $0$	& $0$	& $0$	& $0$	& $0$	& $3$	\\
							
		$\mathbf{v_7}$ 		& $0$	& $0$	& $0$	& $1$	& $2$	& $0$	\\
							
		$\mathbf{v_8}$ 		& $0$	& $0$	& $3$	& $0$	& $0$	& $0$	\\
							
		$\mathbf{v_9}$ 		& $0$	& $0$	& $0$	& $0$	& $2$	& $2$	\\
							
		$\mathbf{v_{10}}$ 	& $3$	& $0$	& $0$	& $0$	& $0$	& $0$	\\
							
		$\mathbf{v_{11}}$ 	& $0$	& $0$	& $0$	& $0$	& $1$	& $1$	\\
							
		$\mathbf{v_{12}}$ 	& $0$	& $0$	& $3$	& $0$	& $0$	& $0$	\\
							
		$\mathbf{v_{13}}$ 	& $2$	& $0$	& $0$	& $0$	& $0$	& $0$	\\
							
		$\mathbf{v_{14}}$ 	& $0$	& $0$	& $0$	& $0$	& $1$	& $0$	\\
							
		$\mathbf{v_{15}}$ 	& $0$	& $1$	& $0$	& $0$	& $0$	& $0$	\\
							
		$\mathbf{v_{16}}$ 	& $0$	& $0$	& $1$	& $0$	& $0$	& $0$	\\
		
		\hline
		\end{tabular}
	\end{threeparttable}
	}
\end{table}


	\begin{table}[t]
	\centering
	\caption{The final information system for clustering threshold $c=0.5$.}\label{table:reduct_infosys2}

	\resizebox{0.8\linewidth}{!}{%
	\begin{threeparttable}
		\begin{tabular}{|ccccccc|l}
		\hhline{-------~}		
		\rowcolor{LightCyan}
					 		& $C_1$ 		& $C_2$ 		& $C_3$ 		& $C_4$
					 		& $C_5$ 		& $C_6$	\\
		\cline{1-7}
		$\mathbf{v_1}$	 	& $1$	& $0$	& $0$	& $0$	& $0$	& $0$
							& \rdelim\}{4}{3mm}[\textbf{Core 1}] 			\\
		$\mathbf{v_5}$ 		& $1$	& $0$	& $0$	& $0$	& $0$	& $0$	\\
		$\mathbf{v_{10}}$	& $1$	& $0$	& $0$	& $0$	& $0$	& $0$	\\
		$\mathbf{v_{13}}$ 	& $1$	& $0$	& $0$	& $0$	& $0$	& $0$	\\
		\cline{1-7}
							
		$\mathbf{v_2}$	 	& $0$	& $0$	& $0$	& $0$	& $0$	& $0$	\\
		\cline{1-7}

		$\mathbf{v_3}$	 	& $0$	& $0$	& $0$	& $0$	& $0$	& $1$
							& \rdelim\}{2}{3mm}[\textbf{Core 2}] 			\\
		$\mathbf{v_6}$ 		& $0$	& $0$	& $0$	& $0$	& $0$	& $1$	\\
		\cline{1-7}

		$\mathbf{v_4}$	 	& $0$	& $0$	& $1$	& $0$	& $0$	& $0$
							& \rdelim\}{4}{3mm}[\textbf{Core 3}] 			\\
		$\mathbf{v_8}$ 		& $0$	& $0$	& $1$	& $0$	& $0$	& $0$	\\
		$\mathbf{v_{12}}$ 	& $0$	& $0$	& $1$	& $0$	& $0$	& $0$	\\
		$\mathbf{v_{16}}$	& $0$	& $0$	& $1$	& $0$	& $0$	& $0$	\\
		\cline{1-7}

		$\mathbf{v_9}$	 	& $0$	& $0$	& $0$	& $0$	& $1$	& $1$
							& \rdelim\}{2}{3mm}[\textbf{Core 4}] 			\\
		$\mathbf{v_{11}}$ 	& $0$	& $0$	& $0$	& $0$	& $1$	& $1$	\\
		\cline{1-7}
		
		$\mathbf{v_7}$	 	& $0$	& $0$	& $0$	& $0$	& $1$	& $0$
							& \rdelim\}{2}{3mm}[\textbf{Core 5}] 			\\
		$\mathbf{v_{14}}$ 	& $0$	& $0$	& $0$	& $0$	& $1$	& $0$	\\
		\cline{1-7}
		
		$\mathbf{v_{15}}$	& $0$	& $1$	& $0$	& $0$	& $0$	& $0$
							& \rdelim\}{1}{3mm}[\textbf{Core 6}] 			\\
		
		\cline{1-7}
		\end{tabular}
	\end{threeparttable}
	}
\end{table}

We observe that cores are built using global clustering information. The final operation is to match pairs of vertices. Cores are visited sequentially one at a time and they are searched locally to find pair matches. Inside each core, a vertex $u$ is selected at random and matched with its ``most similar'' neighbour.  Then the process is repeated with the remaining vertices.  These decisions are made with a local measure of similarity. We use the \textit{Weighted Jaccard Index} that is defined as follows:
\begin{equation}
\label{eq:w_jaccard}
J \left( u,v \right) = \frac{\sum_{\left\lbrace e \triangleright v \, \wedge \, e \triangleright u\right\rbrace} \gamma \left( e \right) }{\sum_{\left\lbrace e \triangleright v \, \vee \, e \triangleright u\right\rbrace} \gamma(e)}, \quad v,u \in V \mbox{, and} \; \forall e \in E.
\end{equation}
This is similar to \textit{non-weighted jaccard index} in \textit{PaToH} (where it is called \emph{Scaled Heavy Connectivity Matching}). This captures similarity in high dimensional datasets better than Euclidean based similarity measures. Vertices that do not find any pair matches during core search are transferred into the non-core vertex list (such as $v_{15}$ in Table~\ref{table:reduct_infosys1}).

One possible problem is that only a small fraction of the hypergraph's vertices might belong to cores. This can happen, for example, the average vertex degree in the hypergraph is high so there is a  large denominator in Eq.(\ref{eq:finalmap}). 

As proposed by Karypis \cite{karytech2002}, we define the \textit{compression ratio} between two levels of coarsening in the multi-level paradigm with $c$ levels as follows:

\begin{equation}\label{eq:comp_ratio}
	r = \frac{\vert V_{i} \vert}{\vert V_{i+1}\vert}, \forall 0 \leqslant i < c
\end{equation}

An advantage of the multi-level approach is that it provides a trade-off between quality and speed-up. The more coarsening levels, the better the partitioning quality, but the longer the running time and the greater the memory consumption. On the other hand, with few coarsening levels, we end up with a large hypergraph in which it is not possible to find a good partitioning.  Having many coarsening levels will give a very small coarsest hypergraph with perhaps few feasible solutions.  This may result in a poor partitioning quality. Karypis \cite{karytech2002} provides a tradeoff between quality and levels of partitioning by limiting the compression ratio $r$ to between $1.5$ and  $1.8$.  In order to satisfy a certain compression ratio, we will visit the non-core vertex list and select vertices randomly. For every selected vertex, the algorithm finds a pair match among its unmatched adjacent vertices in the non-core list.

When pair matches are found, the hypergraph is contracted to build a coarser hypergraph for the next coarsening level. This is done by merging matched vertices. The weight of the coarser vertex is the sum of the weight of two merged vertices and its set of incident hyperedges is the union of the hyperedges incident on each of the merged vertices. After building the coarser hypergraph, we perform two final operations on the hyperedge list. First, hyperedges of unit size are removed as they do not have any impact on the partitioning cut.  Second, identical hyperedges (that is, those having the same vertex set) are detected and each set of identical hyperedges is replaced by a single hyperedge whose weight is the sum of the weight of hyperedges in the set. To find identical hyperedges, each hyperedge is hashed to an integer number based on its vertex list.  To avoid hash conflicts, the content of the hyperedges are compared if they hash to the same number. 

\subsection{Initial Partitioning and Refinement}\label{sec:init_uncoarse}

The coarsening phase ends when the number of the vertices in the coarsest hypergraph is less than a threshold (that is $100$ in our algorithm). The size of the coarsest hypergraph is very small compared to the original hypergraph and its partitioning can be calculated quickly. We use a series of algorithms for this purpose. The best partitioning that preserves the balancing constraint is selected and it is projected back to the original hypergraph. The category of algorithms used for this stage are \textit{random} (randomly assigns vertices to the parts), \textit{linear} (linearly assigns vertices to the parts starting from a random part), and \textit{FM Based} (selects a vertex randomly and assigns it to part 1 and all other vertices to part 0; then the FM algorithm is run and the bipartitioning is developed).

Then partitioning algorithms refine the partitioning as the hypergraph is projected back in the uncoarsening phase. Due to the success of the FM algorithm in practice, we use a variation of FM algorithm known as Early-Exit FM (FM-EE)~\cite{karytech2002} and Boundary FM (BFM)~\cite{ccatalyurek2011patoh}.

\section{Evaluations}\label{sec:evaluations}

In this section we provide the evaluation of our algorithm compared to state-of-the-art partitioning algorithms including \textit{PHG} (the  \textit{Zoltan} hypergraph partitioner) \cite{devetal2006}, \textit{PaToH} \cite{ccatalyurek2011patoh} and \textit{hMetis} \cite{hmetis15url}. \textit{PaToH} is a serial hypergraph partitioner developed by 
\c{C}ataly\"{u}rek 
and it is claimed to be the fastest multilevel recursive bipartitioning based tool. The earliest and the most popular tool for serial hypergraph partitioning is \textit{hMetis} developed by Karypis and Kumar; 
it is specially designed for VLSI circuit partitioning and the algorithms are based on multilevel partitioning schemes and support recursive bisectioning (\textit{shmetis} and  \textit{hmetis}) and direct {$k$-way} partitioning (\textit{kmetis}). Finally, \textit{Zoltan} data management services for parallel dynamic applications is a toolkit developed at Sandia National Laboratories. The library includes a wide range of tools for problems such as Dynamic Load Balancing, Graph/Hypergraph Colouring, Matrix Operations, Data Migration, Unstructured Communications, Distributed Directories and Graph/Hypergraph Partitioning. It follows a distributed memory model and uses MPI for interprocessor communications. It is available within \textit{Trilinos}, an open source software project for scientific applications, since version 9.0 \cite{trilinosurl}.  

All of these algorithms are multi-level recursive bipartitioning algorithms and \textit{PHG} is a parallel hypergraph partitioner while the other two are serial partitioning tools.

For the evaluation, we have selected a number of test hypergraphs from a variety of  scientific applications with different specifications. The hypergraphs are obtained from the University of Florida Sparse Matrix Collection \cite{sparsecollection2011}. It is a large database of sparse matrices from real applications. Each sparse matrix from the database is treated as the hypergraph incident matrix with the vertices and hyperedges as rows and columns of the matrix, respectively. This is similar to the column-net model proposed in \cite{catayk1999}. The weight of each vertex and each hyperedge is assumed to be 1. The list of test data used for our evaluation is in Table \ref{tab:fehg_testdata}. 

\begin{table*}[t]
	\centering
	\captionsetup{font=small,labelfont=bf}
	\caption{Tested hypergraphs for sequential algorithm simulation and their specifications} \label{tab:fehg_testdata}
	\resizebox{0.85\textwidth}{!}{%
	\begin{threeparttable}
		\begin{tabular}{|l|c|c|c|c|c|}
		\hline
		Hypergraph & Description & Rows & Columns & Non-Zeros & NSCC\tnote{1}\; \\
		\hline

		CNR-2000 		
& Small web crawl of Italian CNR domain	& 325,557	& 325,557 	& 3,216,152 &	100,977	\\

		AS-22JULY06		
& Internet routers					  	& 22,963		& 22,963		& 96,872		& 1	\\

		CELEGANSNEURAL	
& Neural Network of Nematode C. Elegans	& 297		& 297		& 2,345		& 57	\\

		NETSCIENCE		
& Co-authorship of scientists in Network Theory & 1,589 & 1,589		& 5,484	& 396	\\

		PGPGIANTCOMPO	
& Largest connected component in graph of PGP users 	& 10,680 & 10,680 & 48,632	& 1	\\

		GUPTA1			
& Linear Programming matrix ($A \times A^T$)	& 31,802	& 31,802		& 2,164,210		& 1 \\

		MARK3JAC120		
& Jacobian from MULTIMOD Mark3			& 54,929		& 54,929		& 322,483	&	1,921 \\

		NOTREDAME\_WWW
& Barabasi's web page network of nd.edu	& 325,729	& 325,729	& 929,849	&	231,666\\

		PATENTS\_MAIN	
& Pajek network: mainNBER US Patent Citations & 240,547 & 240,547 & 560,943	&	240,547 \\

		STD1\_JAC3		
& Chemical process simulation			& 21,982		& 21,982		& 1,455,374	& 1	\\

		COND-MAT-2005	
& Collaboration network, www.arxiv.org	& 40,421		& 40,421		& 351,382	& 1,798	\\
		\hline
		\end{tabular}
		\begin{tablenotes}
			\small
			\item[1] NSCC stands for the Number of Strongly Connected Components.
		\end{tablenotes}
	\end{threeparttable}
	} 
\end{table*}

The simulations are done on a computer with Intel(R) Xeon(R) CPU E5-2650 2.00GHz processor, 8GB of RAM and 40GB of disk space and the operating system running on the system is 32-Bit Ubuntu 12.04 LTS. Furthermore, we set the imbalance tolerance to $2\%$ and the number of parts (the value of $k$) is each of $\left\lbrace 2,4,8,16,32 \right\rbrace$. The final imbalances achieved by the algorithms are not reported because the balancing requirement was always met.

\subsection{Algorithm Parameters}\label{sec:fehg_parameters}

Each of the evaluated tools has different input parameters that can be set by the user. We use default parameters for each tool. All algorithms use a variation of the FM algorithm (FM-EE and BFM) in their refinement phase. \textit{PHG} uses an \textit{agglomerative} coarsening algorithm that uses the \textit{inner product} as the measure of similarity between the vertices \cite{devetal2006}. This is also used in \textit{hMetis}. The default partitioning tool for \textit{hMetis} is \textit{shmetis} and the default coarsening scheme is the \textit{Hybrid First Choice (HFC)} scheme which is a combination of the First Choice (FC) \cite{karytech2002} and Greedy First Choice scheme (which is a variation of the FC algorithm in which vertices are grouped and the grouping is biased in favor of faster reduction in the number of the hyperedges in the coarser hypergraph). \textit{PaToH} is initialised by setting the \textsf{SBProbType} parameter to \textsf{PATOH\_SUGPARAM\_DEFAULT}. It uses Absorption Clustering using Pins as the default coarsening algorithm (this is an agglomerative vertex clustering scheme). The similarity metric, known as \textbf{absorption metric}, of two vertices $u$ and $v$ is calculated as follows:
\begin{equation}
\sum_{\left\lbrace \forall e \in E \mid u \in e \textit{ and } v \in e \right\rbrace } \frac{1}{\vert e \vert - 1}.
\end{equation}

The algorithm finds the absorption metric for every pin that connects $u$ and the cluster that vertex $v$ is already assigned to. See the manuals of the tools for the full description of the parameters \cite{ccatalyurek2011patoh,karypis1998hmetis,zoltanurl}.

The \textit{FEHG} algorithm has two parameters that need to be set: the \textit{similarity threshold} in Definition \ref{def:hcg} for building the HCG, and the clustering threshold in Eq.(\ref{eq:finalmap}). In the rest of this subsection, we look further at these two parameters. 

In graph theory, a set of vertices are said, informally, to be \textit{clustered} if they induce more edges (or hyperedges) than one would expect if the edges had been placed in the graph at random.  There are various ways in which one can formally define a clustering coefficient (CC) of either a vertex or a graph, and for hypergraphs various definitions have been proposed \cite{gavin2002functional,klamt2009hypergraphs,latapy2008basic}.  We need something a little different as we are interested in the clustering of hyperedges and we want to take account of their weights.  Given a hypergraph $H=(V,E)$, we define {CC} for a hyperedge $e \in E$ as follows:

\begin{dmath}\label{eq:cc}
\text{CC}(e) =
\begin{cases}
 \frac{\sum\nolimits_{\left\lbrace e' \cap e \neq \emptyset \right\rbrace} \left( \left( \frac{\left\vert {e \cap e'} \right\vert}{\left\vert {e} \right\vert-1} \right) \cdot \gamma(e') \right)}{\sum\nolimits_{\left\lbrace v \in e \right\rbrace} \sum\nolimits_{\left\lbrace e'' \triangleright v \right\rbrace } \gamma(e'')}, \, \forall e',e'' \in E \backslash e, & \mbox{if} \, \left\vert {e} \right\vert >1 \\
 0, & \mbox{otherwise},
\end{cases}
\end{dmath}
and the CC of $H$ is calculated as the average of CC over all its hyperedges: 

\begin{equation}\label{eq:cc_hgraph}
{CC}_{H} = \sum_{e \in E} {\frac{CC(e)}{\vert E \vert}}.
\end{equation}

During coarsening, each time we proceed to the next coarsening level, the structure of the hypergraph changes and so the value of CC in the coarser hypergraph is different. To recalculate the CC value in  each coarsening level could be costly so we are interested in finding a way of \textit{updating} the CC value without a complete recalculation (possibly sacrificing accuracy in the process). Foudalis et al. \cite{foudalis2011social} studied the structure of social network graphs and looked at several characteristic metrics including the clustering coefficient. They found that social networks have high CC compared to random networks, and that the CC is negatively correlated to the degree of vertices. Two vertices with low vertex degree are more likely to cluster to each other than two vertices with higher degree. In addition, Bloznelis \cite{bloznelis2013} has theoretically investigated random intersection graphs\footnote{Random intersection graphs can be obtained from randomly generated bipartite graphs which have bipartition $V \cup W$ if each vertex $v_i$ in $V=\left\lbrace v_1, v_2, \cdots v_n \right\rbrace$ selects the set $D_i \subset W$ of its neighbours in the bipartite graph randomly and independently such that the elements of $W$ have equal probability to be selected. This can be considered to define a hypergraph by assuming that $V$ and $W$ are the sets of vertices and hyperedges, respectively.} and shown that the CC is inversely related to the average vertex degree in the graph. Based on these results, we update the value of CC from one coarsening level to the next based on the inverse of the average vertex degree. Finally, the similarity threshold is set as the CC of the hypergraph at the beginning of the coarsening phase. In the simulation section, we investigate how our algorithm performs, in terms of partition quality and running time, when the similarity threshold is updated compared to when it is recalculated in each coarsening level.

\begin{figure}[t]
	\captionsetup{font=small,labelfont=bf}
	\centering
	\resizebox {0.45\textwidth} {!} {
	\fbox{\includegraphics{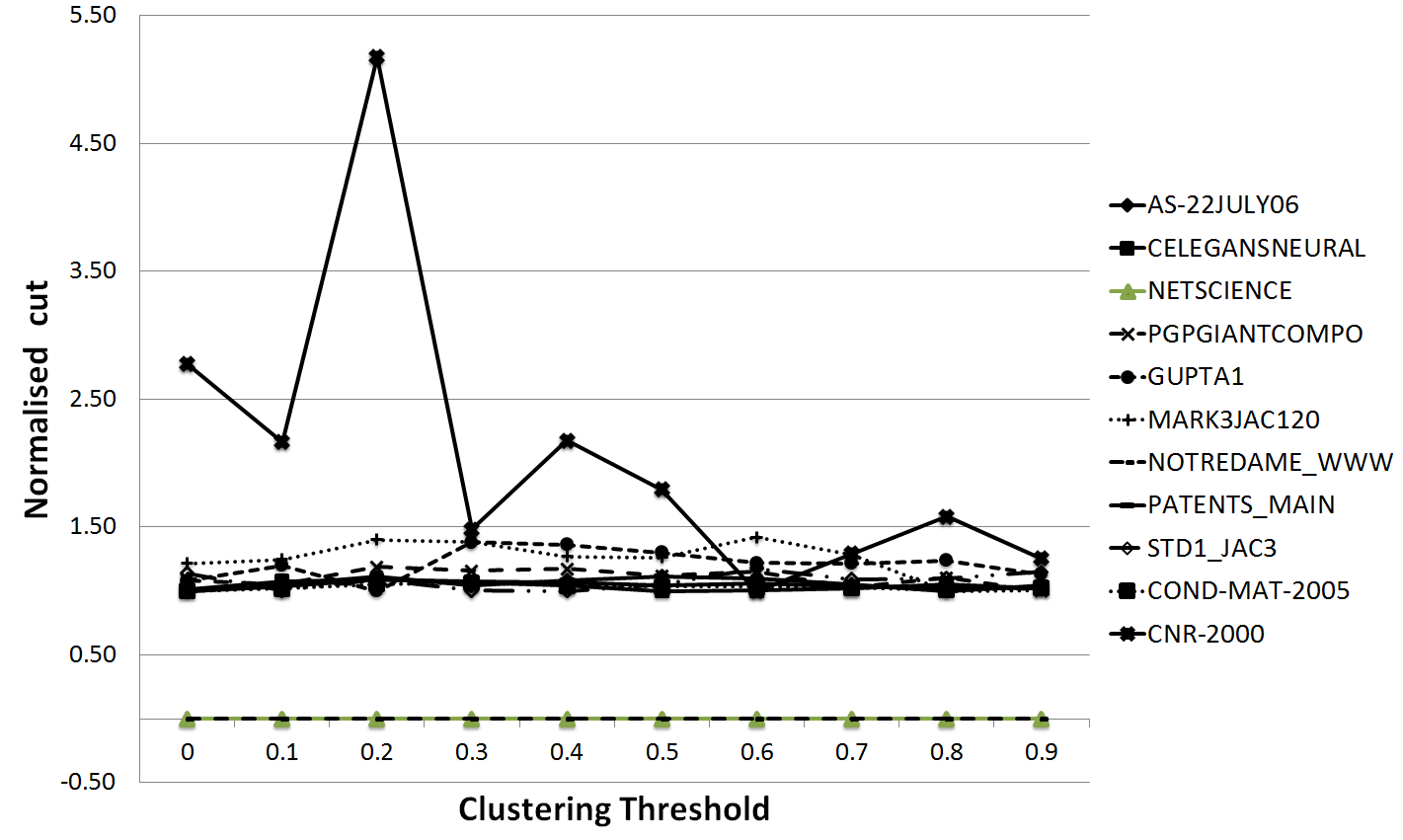}}
	}
\caption{The variation of bipartitioning cut based on the clustering threshold. Values are normalised with the best cut for each hypergraph.}\label{fig:clus_thre_var}
\end{figure}

In order to investigate the impact of the clustering threshold on the quality of the partitioning, we have checked the dependency of the algorithm on the clustering threshold. Figure~\ref{fig:clus_thre_var} depicts the quality of the {2-way} partitioning of the hypergraphs for variable clustering threshold values. The cut is normalised based on the best partitioning cut for each hypergraph. The average correlation between the cut and the clustering threshold over all hypergraphs is calculated to be $0.2442$ and excluding \texttt{CNR-2000} is $0.2096$, a very weak correlation. The standard deviation of the changes with respect to the average cut values is also less than $4.2\%$ over all hypergraphs. Therefore, changing the clustering threshold has a very small effect on the partitioning quality. (An exception occurs for the \texttt{CNR-2000} hypergraph such that the variation of change is very high.) Hyperedges with high CC values are those that are more likely clustered with other hyperedges and those with low CC values do not form any cluster and form edge partitions in HCG of size 1. Consequently, we can remove every edge partition in HCG of unit size and set the value of the clustering threshold to zero in Eq.(\ref{eq:finalmap}). As an example, edge partitions $C_2$ and $C_4$ can be removed from the reduced information system in Table~\ref{table:reduct_infosys2} without causing any changes in the cores. Using this strategy, the partitioning cut for \texttt{CNR-2000} is $29.9\%$ better than the best bipartitioning cut reported in Fig.~\ref{fig:clus_thre_var} and this is achieved for $c=0.6$.

\subsection{Simulations}\label{sec:simulations}

In the first evaluation, we assume unit weight for both vertices and hyperedges of the hypergraph. In this situation, a partitioning algorithms performs well if it can capture strongly connected components of the hypergraph.  A strongly connected component is a group of the vertices that are tightly coupled together. Assuming that the hypergraph represents a graph (by limiting the hyperedge cardinality to two), a strongly connected component is a clique. Because the weight of all hyperedges is 1, the aim of the partitioning algorithm is to take these cliques out of the cut as they are the major cause of increasing the cut; therefore, the vertex connectivity is more important for identifying those cliques. An algorithm that identifies those cliques and merged their vertices to build a coarser vertex is the one that gives better partitioning quality and the clustering algorithm that captures those strongly connected components in the first few levels of coarsening would obtain competitive partitioning qualities.

\setlength{\extrarowheight}{3pt}

\begin{table*}[hp]
	\captionsetup{font=small,labelfont=bf}
	\caption{Quality comparison of the algorithms for different part sizes and $2\%$  imbalance tolerance. The values are normalised according to the minimum value for each hypergraph; therefore, the algorithm that gives $1.0$ cut value is considered to be the best. Unit weights are assumed for both vertices and hyperedges.}\label{tab:unit_weights_comp_alg}
	\centering
	\resizebox{0.9\textwidth}{!}{%
	\begin{threeparttable}
		\begin{tabular}{|l|c|cc|cc|cc|cc|cc|}
			\hline
			& 	&\multicolumn{10}{|c|}{} \\
			& 	&\multicolumn{10}{c|}{\textbf{Number of Parts}} \\
			& 	& \multicolumn{2}{c}{2} & \multicolumn{2}{c}{4} 
				& \multicolumn{2}{c}{8} & \multicolumn{2}{c}{16} 
				& \multicolumn{2}{c|}{32} \\
			\cline{3-12} 
			& 				& \textbf{AVE} & \textbf{BEST}
							& \textbf{AVE} & \textbf{BEST}
							& \textbf{AVE} & \textbf{BEST}
							& \textbf{AVE} & \textbf{BEST}
							& \textbf{AVE} & \textbf{BEST}
			\\
			\hline

			& FEHG			& 1.11	& 1.00	& 1.02	& 1.00	& 1.04	& 1.01	& 1.01	& 1.00	& 1.01	& 1.03 \\
\textbf{ AS-22JULY06}
			& PHG			& 2.90	& 2.46	& 1.77	& 1.56	& 1.64	& 1.36	& 1.43	& 1.34	& 1.37	& 1.32 \\
			& hMetis			& 1.34	& 1.95	& 1.19	& 1.30	& 1.16	& 1.18	& 1.04	& 1.06	& 1.09	& 1.04  \\
			& PaToH			& 1.00	& 1.43	& 1.00	& 1.03	& 1.00	& 1.00	& 1.00	& 1.00	& 1.00	& 1.00 \\
			\cline{2-12}
			& Min Value 	& 136 	& 93		& 355	& 319	& 629	& 599 	& 1051	& 995	& 1591	& 1529 \\
			\hline
			& FEHG			& 1.00	& 1.00	& 1.09	& 1.00	& 1.10	& 1.06	& 1.11	& 1.08	& 1.07	& 1.03 \\
\textbf{ CELEGANSNEURAL}
			& PHG			& 1.07	& 1.00	& 1.04	& 1.03	& 1.02	& 1.00	& 1.06	& 1.00	& 1.00	& 1.00 \\
			& hMetis			& 1.17	& 1.21	& 1.00	& 1.05	& 1.00	& 1.04	& 1.00	& 1.02	& 1.00	& 1.00 \\
			& PaToH			& 1.01	& 1.04	& 1.00	& 1.06	& 1.03	& 1.07	& 1.03	& 1.06	& 1.05	& 1.05 \\
			\cline{2-12}
			& Min Value 	& 79& 77	& 195	& 184	& 354	& 342	& 548	& 536	& 773	& 769 \\
			\hline

			& FEHG			& 1.37	& 1.00	& 1.71	& 1.07	& 1.59	& 1.41	& 1.53	& 1.45	& 1.63	& 1.51 \\
\textbf{ CNR--2000}
			& PHG			& 35.88	& 45.62	& 12.48	& 9.17	& 5.73	& 4.84	& 3.54	& 2.98	& 2.42	& 2.02 \\
			& hMetis			& 12.19	& 18.82	& 8.24	& 8.43	& 5.08	& 4.71	& 3.46	& 3.29	& 2.66	& 2.50 \\
			& PaToH			& 1.00	& 1.71	& 1.00	& 1.00	& 1.00	& 1.00	& 1.00	& 1.00	& 1.00	& 1.00 \\
			\cline{2-12}
			& Min Value 	& 81		& 45		& 244	& 202	& 569	& 509	& 1014	& 911	& 1927& 1830\\
			\hline

			& FEHG			& 1.00	& 1.00	& 1.00	& 1.00	& 1.00	& 1.00	& 1.01	& 1.02	& 1.01	& 1.00 \\
\textbf{ COND--MAT--2005}
			& PHG			& 1.17	& 1.17	& 1.11	& 1.10	& 1.05	& 1.05	& 1.03	& 1.03	& 1.02	& 1.01 \\
			& hMetis			& 1.05	& 1.07	& 1.11	& 1.12	& 1.11	& 1.12	& 1.11	& 1.10	& 1.01	& 1.01 \\
			& PaToH			& 1.02	& 1.02	& 1.03	& 1.03	& 1.00	& 1.00	& 1.00	& 1.10	& 1.00	& 1.00 \\
			\cline{2-12}
			& Min Value		& 2134	& 2087	& 5057	& 4951	& 8609	& 8485	& 12370	& 12150	& 16270 & 16150 \\
			\hline

			& FEHG			& 0.0	& 0.0	& 0.0	& 0.0	& 2.00	& 1.50	& 1.50	& 1.00	& 2.08	& 1.81 \\
\textbf{ NETSCIENCE\tnote{\textasteriskcentered}}
			& PHG			& 0.0	& 0.0	& 0.0	& 0.0	& 1.50	& 1.00	& 1.40	& 1.00	& 1.87	& 1.5 \\
			& hMetis			& 2.0	& 2.0	& 5.0	& 5.0	& 4.22	& 3.50	& 1.75	& 1.75	& 1.99	& 1.87 \\
			& PaToH			& 0.0	& 0.0	& 0.0	& 0.0	& 1.00	& 1.00	& 1.00	& 1.00	& 1.00	& 1.00 \\
			\cline{2-12}
			& Min Value 	& 0	& 0		& 0		& 0		& 2		& 2		& 8	& 8	& 16		& 16	 \\
			\hline

			& FEHG			& 2.12	& 1.27	& 1.00	& 1.00	& 1.04	& 1.00	& 1.00	& 1.08	& 1.00	& 1.00 \\
\textbf{ PGPGIANTCOMPO}
			& PHG			& 13.23	& 1.83	& 1.44	& 1.04	& 1.25	& 1.04	& 1.02	& 1.00	& 1.08	& 1.00	 \\
			& hMetis			& 9.7	& 9.61	& 1.46	& 1.71	& 1.04	& 1.40	& 1.31	& 1.40	& 1.26	& 1.27 \\
			& PaToH			& 1.00	& 1.00	& 1.04	& 1.27	& 1.00	& 1.04	& 1.02	& 1.15	& 1.08	& 1.06 \\
			\cline{2-12}
			& Min Value 	& 18		& 18		& 242	& 200	& 419	& 400	& 695	& 617	& 956	& 930 \\
			\hline

			& FEHG			& 1.00	& 1.00	& 1.00	& 1.00	& 1.00	& 1.00	& 1.00	& 1.00	& 1.00	& 1.00 \\
\textbf{ GUPTA1}
			& PHG			& 1.58	& 1.45	& 1.31	& 1.24	& 1.15	& 1.04	& 1.07	& 1.04	& 1.09	& 1.05 \\
			& hMetis			& 1.73	& 1.82	& 1.61	& 1.69	& 1.58	& 1.64	& 1.60	& 1.57	& 1.51	& 1.48 \\
			& PaToH			& 1.22	& 1.17	& 1.08	& 1.09	& 1.04	& 1.05	& 1.05	& 1.07	& 1.08	& 1.09 \\
			\cline{2-12}
			& Min Value		& 486	& 462	& 1466	& 1384	& 3077	& 2893	& 5342	& 5134	& 8965& 8519 \\
			\hline

			& FEHG			& 1.01	& 1.01	& 1.02	& 1.01	& 1.01	& 1.00	& 1.00	& 1.00	& 1.06	& 1.07 \\
\textbf{ MARK3JAC120}
			& PHG			& 1.00	& 1.01	& 1.02	& 1.02	& 1.02	& 1.00	& 1.00	& 1.00	& 1.72	& 1.78 \\
			& hMetis			& 1.00	& 1.00	& 1.00	& 1.02	& 1.00	& 1.00	& 1.30	& 1.00	& 4.20	& 1.78 \\
			& PaToH			& 1.00	& 1.02	& 1.00	& 1.00	& 1.00	& 1.00	& 1.26	& 1.20	& 1.00 	& 1.00 \\
			\cline{2-12}
			& Min Value		& 408	& 400	& 1229	& 1202	& 2856	& 2835	& 6317	& 6245	& 3142& 2944 \\
			\hline

			& FEHG			& 0		& 0		& 1.00	& 1.00	& 1.12	& 1.12	& 1.09	& 1.03	& 1.06	& 1.07 \\
\textbf{ NOTREDAME\tnote{\textasteriskcentered}}
			& PHG			& 4326	& 4326	& 158.56	& 288.69& 13.82	& 16.78& 2.09	& 3.06	& 1.72	& 1.78 \\
			& hMetis			& 880	& 707	& 67.92	& 129.92& 10.98	& 12.65& 3.36	& 3.37	& 2.23	& 2.30 \\
			& Patoh			& 24		& 22		& 1.90	& 3.31	& 1.00	& 1.00	& 1.00	& 1.00	& 1.00	& 1.00 \\
			\cline{2-12}
			& Min Value		& 0		& 0		& 27		& 13		& 316	& 259	& 1577	& 1484	& 3142 & 2944 \\
			\hline

			& FEHG			& 1.20	& 1.00	& 1.03	& 1.01	& 1.05	& 1.03	& 1.00	& 1.00	& 1.00	& 1.00 \\
\textbf{ PATENTS--MAIN}
			& PHG			& 12.49	& 13.19	& 2.52	& 2.30	& 1.79	& 1.65	& 1.42	& 1.38	& 1.23	& 1.18 \\
			& hMetis			& 2.38	& 2.77	& 1.16	& 1.24	& 1.26	& 1.43	& 1.26	& 1.31	& 1.21	& 1.22 \\
			& PaToH			& 1.00	& 1.02	& 1.00	& 1.00	& 1.00	& 1.00	& 1.00	& 1.00	& 1.01	& 1.00 \\
			\cline{2-12}
			& Min Value		& 643	& 528	& 3490	& 3198	& 6451	& 6096	& 11322	& 10640	&16927 & 16460 \\
			\hline

			& FEHG			& 1.01	& 1.00	& 1.00	& 1.03	& 1.00	& 1.00	& 1.00	& 1.00	& 1.00	& 1.00 \\
\textbf{ STD1--JAC3}
			& PHG			& 1.15	& 1.08	& 1.16	& 1.10	& 1.18	& 1.13	& 1.28	& 1.35	& 1.33	& 1.29 \\
			& hMetis			& 1.05	& 1.00	& 1.52	& 1.03	& 1.54	& 1.23	& 1.70	& 1.53	& 1.71	& 1.51 \\
			& Patoh			& 1.00	& 1.00	& 1.08	& 1.00	& 1.16	& 1.14	& 1.00	& 1.26	& 1.30	& 1.29 \\
			\cline{2-12}
			& Min Value		& 1490	& 1371	& 3735	& 3333	& 7616	& 6167	& 13254	& 11710	& 22242 & 21200\\

			\hline
		\end{tabular}
		\begin{tablenotes}
			\item[\textasteriskcentered] When the minimum cut for the average or best cases are zero, the values
			 shown are actual cut values rather than normalised values. 
		\end{tablenotes}
	\end{threeparttable}
	}	
\end{table*}

Each algorithm is run 20 times and its average cut is stated in Table~\ref{tab:unit_weights_comp_alg} as well as the best cut among all runs. The results in the table are normalised with respect to the minimum value among algorithms. For example, \textit{FEHG} finds the minimum average cut of 79 for a bipartitioning on the \texttt{CELEGANSNEURAL} hypergraph. \textit{PHG}, \textit{hMetis} and \textit{PaToH}  find, respectively, average bipartitioning cuts that are worse by factors of $1.07$, $1.17$, and $1.01$.  The results show that \textit{FEHG} performs very well compared to \textit{PHG} and \textit{hMetis} and it is competitive with \textit{PaToH}; considering the best cut, this is found in 30 (of the 55 cases) by \textit{FEHG} and in 27 of the cases by \textit{PaToH} (this includes some ties). As can be seen from the results, all algorithms give similar partitioning cut when the hypergraph has only a few strongly connected components. In this situation, even the local clustering algorithms can capture these strongly connected components and merge their vertices; therefore the differences in partitioning cut for different algorithms that are using different clustering methods (either global or local) is very small.

As the number of strongly connected components increases, it is harder to identify those components especially when the boundaries between them are not clear and they have overlaps. This situation happens in hypergraphs such as \texttt{Notredame}, \texttt{Patents-Main}, and \texttt{CNR-2000}. As shown, \textit{FEHG} achieves a superior quality improvement compared to Zoltan and hMetis, but \textit{PaToH} still generates very good partitionings with absorption clustering using pins. One reason that may explain this is that \textit{PaToH} allows matching between a group of vertices instead of pair-matching. Therefore, the algorithm can merge strongly connected components of vertices in the first few levels of coarsening and remove them from the cut. Though \textit{hMetis} also allows multiple matches, it seems that the agglomerative clustering strategy of \textit{PaToH} is much better compared to the hybrid first choice algorithm in \textit{hMetis}. 

The standard deviation (STD) from the average cut is calculated for some of the algorithms and is reported in Table~\ref{tab:unit_weights_std}. This shows the reliability of the partitioning algorithm when it is used in practical applications. The values are reported as a percentage of STD based on the average partitioning cut for each algorithm. We note, first, that the standard deviation is an increasing function of the number of parts (this is due to the recursive bipartitioning nature of the algorithms which adds to the STD in each recursion), and, second, that the percentage of the STD based on the average cut is decreasing as the number of parts increases (as the average cut increases exponentially when as the number of parts increases, while the increase of the STD is linear). We observe very good reliability for \textit{hMetis} despite the fact that it gives the worst quality (on average) compared to the others. If we consider bring the average of the cut into the calculation, the least reliability is obtained for \textit{hMetis} and \textit{PHG}. Our evaluations recognises \textit{FEHG} and \textit{PHG} as the most reliable algorithms compared to others. 

\begin{table}[ht]
	\captionsetup{font=small,labelfont=bf}
	\caption{The percentage of the Standard Deviation (STD) of the cut based on the average partitioning cut for each algorithm with variable number of parts. Unit weights are assumed for both vertices and hyperedges.}\label{tab:unit_weights_std}
	\centering
	\small
	\resizebox{0.45\textwidth}{!}{%
	\begin{threeparttable}
		\begin{tabular}{|l|c|ccccc|}
			\hline
			& &\multicolumn{5}{|c|}{\textbf{Number of Parts}} \\ 
			\cline{3-7}
			& & \textbf{2} & \textbf{4} & \textbf{8} & \textbf{16} & \textbf{32}\\
			\hline
			
			& FEHG	& 22.4 & 8.8 & 3.8 & 2.8 & 1.7 \\
AS-22JULY06
			& PHG	 & 21.8 & 14.6 & 7.6 & 5.8 & 4.1 \\
			& hMetis	 & 0 & 1.6 & 1.6 & 2.1 & 1.6 \\
			& PaToH	& 2.9 & 4.5 & 3.2 & 3.5 & 2.7\\
			\hline

			& FEHG	& 56.4 & 31.3 & 24.8 & 14 & 6.9 \\
CNR-2000
			& PHG	& 18.9 & 24.9 & 17.4 & 13.3 & 11.4 \\
			& hMetis	& 7.5 & 8.1 & 8.3 & 6.8 & 4.5 \\
			& PaToH	& 47.9 & 79 & 17.2 & 15.1 & 9.2 \\
			\hline

			& FEHG	& 1.3 & 1.1 & 1.0 & 0.7 & 0.6\\
COND\_MAT
			& PHG	& 1.5 & 1.5 & 1 & 0.9 & 0.8 \\
			& hMetis& 0.6 & 1.3 & 0.8 & 0.9 & 0.8 \\
			& PaToH	& 1.8 & 3.7 & 1.1 & 1.2 & 1.1\\
			\hline

			& FEHG	& 8 & 23 & 18 & 16 & 18 \\
PGPGIANT
			& PHG	& 48 & 65 & 45 & 53 & 46 \\
			& hMetis& 3 & 11 & 13 & 24 & 25 \\
			& PaToH	& 0 & 0 & 7 & 2 & 5\\
			\hline

			& FEHG	& 20 & 9.5 & 4.1 & 2.3 & 1.9 \\
GUPTA1
			& PHG	& 20.1 & 18.6 & 8.6 & 7.5 & 4.4 \\
			& hMetis	& 1.7 & 3.1 & 2.2 & 2.6 & 2.1 \\
			& PaToH	& 0 & 0 & 1.7 & 0.3 & 0.5 \\
			\hline

			& FEHG	& 1.5 & 1.4 & 0.8 & 1.3 & 3.9 \\
MARK3JAC
			& PHG	& 1 & 1.2 & 0.9 & 0.8 & 2.0 \\
			& hMetis& 3.2 & 1.2 & 1 & 2.6 & 1.6 \\
			& PaToH	& 0 & 0.9 & 0.6 & 3.1 & 8.5 \\
			\hline

			& FEHG	& 0 & 33.2 & 11.2 & 6.7 & 3.6 \\
NOTREDAME
			& PHG	& 0 & 2.9 & 1.5 & 1.6 & 1.4 \\
			& hMetis	& 9.5 & 3.5 & 3.1 & 2.7 & 1.8 \\
			& Patoh	& 4.1 & 15.5 & 8.5 & 3.3 & 2 \\
			\hline

			& FEHG	& 23.2 & 7.6 & 4 & 2.9 & 2 \\
PATENTS
			& PHG	& 16 & 19.7 & 15.1 & 9.8 & 7.7 \\
			& hMetis	& 2.4 & 1.7 & 1.4 & 1.1 & 1.1 \\
			& PaToH	& 10.9 & 4.2 & 3.4 & 2 & 1.8 \\
			\hline

			& FEHG	& 17.3 & 6.6 & 5.6 & 4.1 & 2.5 \\
STD1\_JAC3
			& PHG	& 13.3 & 8.7 & 8.3 & 4.5 & 2.7 \\
			& hMetis	& 6.7 & 29.1 & 17.5 & 10.3 & 7.9 \\
			& Patoh	& 8.4 & 12.6 & 7.9 & 6.2 & 3.3 \\

			\hline
		\end{tabular}
	\end{threeparttable}
	}	
\end{table}

\begin{figure}[t]
	\centering
	\begin{subfigure}{0.45\textwidth}
		\centering
		\captionsetup{font=small,labelfont=bf}
		\resizebox {\textwidth} {!} {
		\fbox{\includegraphics{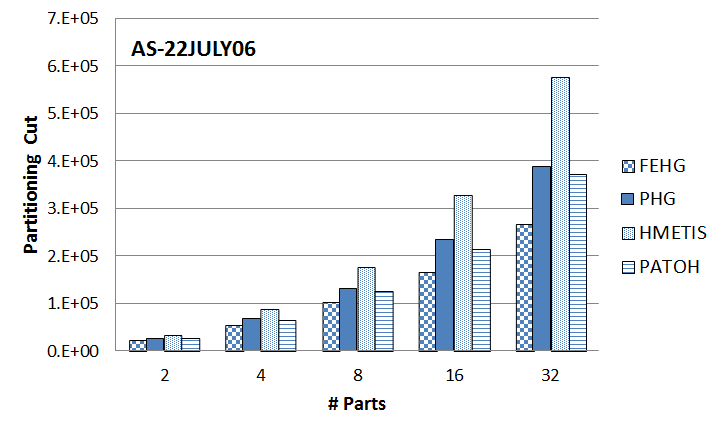}}
		}
	\caption{}
	\end{subfigure}
	
	\begin{subfigure}{0.45\textwidth}
		\centering
		\captionsetup{font=small,labelfont=bf}
		\resizebox {\textwidth} {!} {
		\fbox{\includegraphics{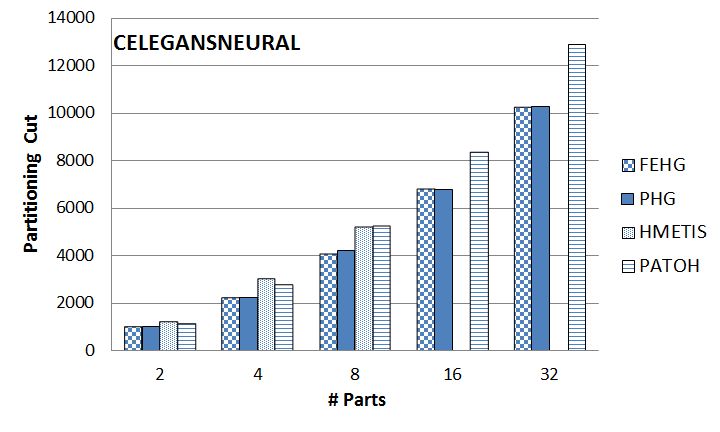}}
	}
	\caption{No results for \textit{hMetis} for 16 and 32 parts.}
	\end{subfigure}

	\begin{subfigure}{0.45\textwidth}
		\centering
		\captionsetup{font=small,labelfont=bf}
		\resizebox {\textwidth} {!} {
		\fbox{\includegraphics{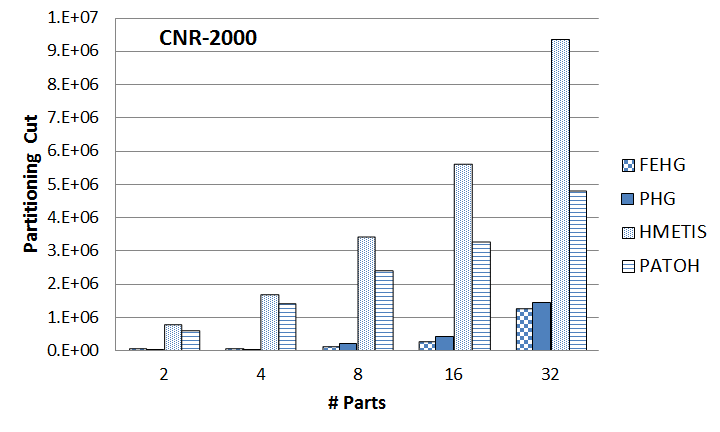}}
	}
	\caption{}
	\end{subfigure}
		
	\caption{Comparing the variation of the average cut for different partitioning numbers. The weight of vertices are unit and the weight of hyperedges are their sizes.}\label{fig:partcut_scale_edge}
\end{figure}	

\begin{figure}[t]
	\centering
    \ContinuedFloat 

	\begin{subfigure}{0.45\textwidth}
		\centering
		\captionsetup{font=small,labelfont=bf}
		\resizebox {\textwidth} {!} {
		\fbox{\includegraphics{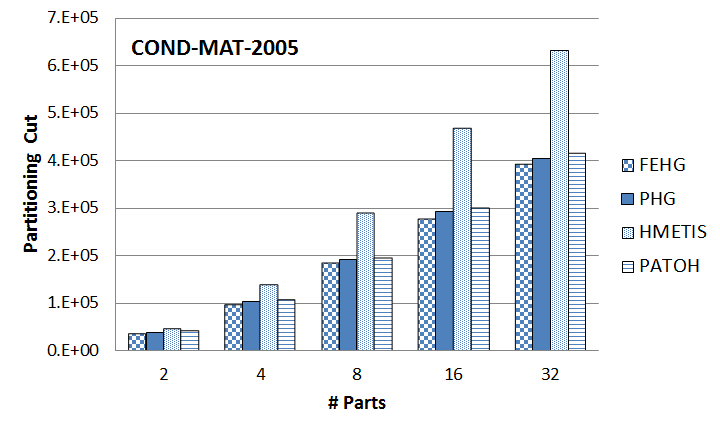}}
		}
	\caption{}
	\end{subfigure}
	
	\begin{subfigure}{0.45\textwidth}
		\centering
		\captionsetup{font=small,labelfont=bf}
		\resizebox {\textwidth} {!} {
		\fbox{\includegraphics{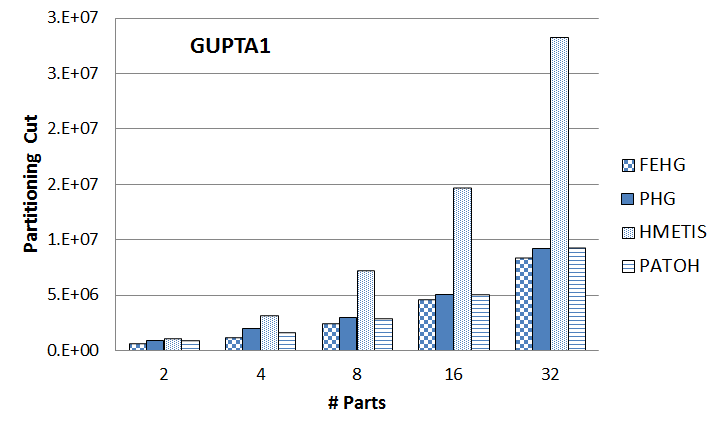}}
	}
	\caption{}
	\end{subfigure}

	\begin{subfigure}{0.45\textwidth}
		\centering
		\captionsetup{font=small,labelfont=bf}
		\resizebox {\textwidth} {!} {
		\fbox{\includegraphics{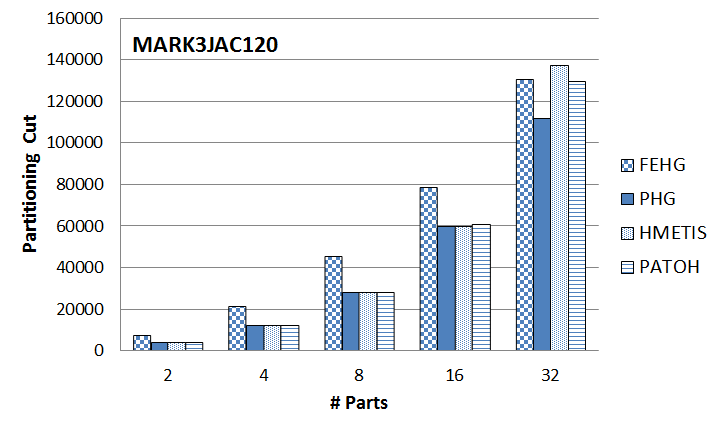}}
	}
	\caption{}
	\end{subfigure}

\caption{(Continued) Comparing the variation of the average cut for different partitioning numbers. The weight of vertices are unit and the weight of hyperedges are their sizes.}
\end{figure}


\begin{figure}[t]
	\centering
    \ContinuedFloat 

	\begin{subfigure}{0.45\textwidth}
		\centering
		\captionsetup{font=small,labelfont=bf}
		\resizebox {\textwidth} {!} {
		\fbox{\includegraphics{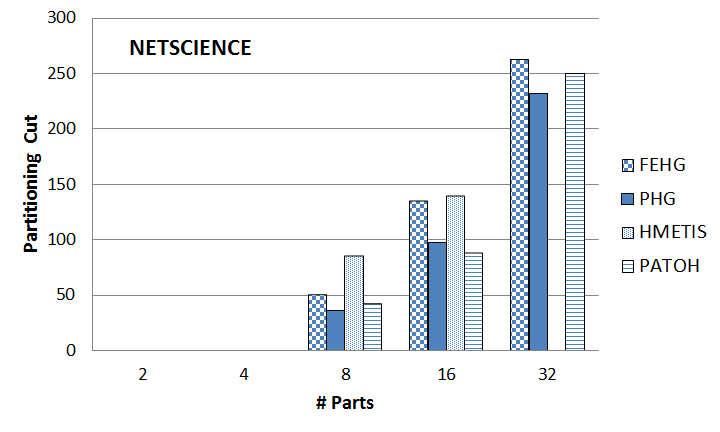}}
		}
	\caption{Cut of size zero found for {2-way} and {4-way} partitioning. No results for {hMetis} for \textit{32-way} partitioning.}
	\end{subfigure}

	\begin{subfigure}{0.45\textwidth}
		\centering
		\captionsetup{font=small,labelfont=bf}
		\resizebox {\textwidth} {!} {
		\fbox{\includegraphics{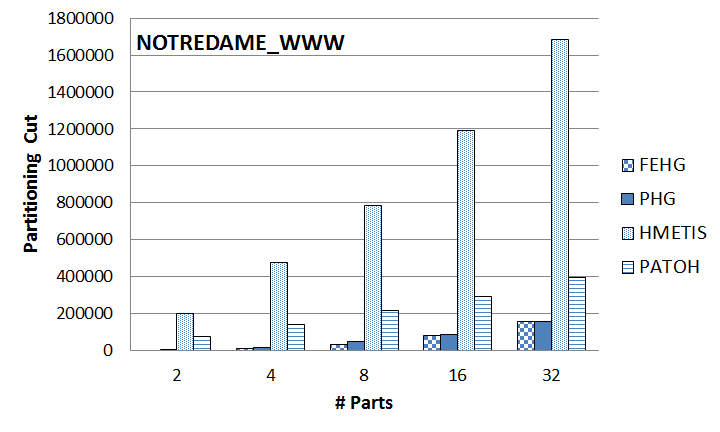}}
	}
	\caption{}
	\end{subfigure}

	\begin{subfigure}{0.45\textwidth}
		\centering
		\captionsetup{font=small,labelfont=bf}
		\resizebox {\textwidth} {!} {
		\fbox{\includegraphics{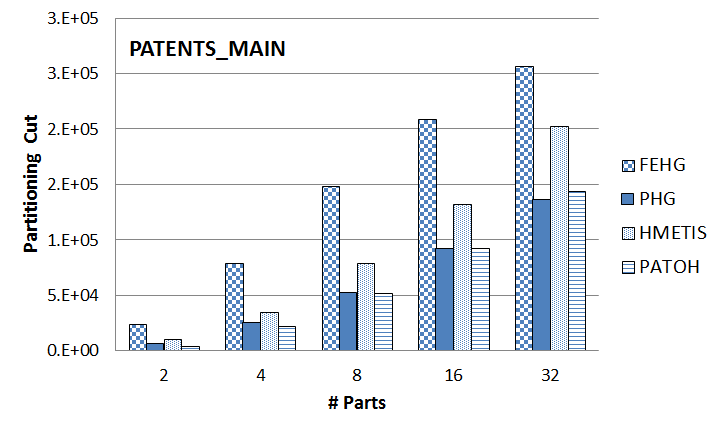}}
	}
	\caption{}
	\end{subfigure}

\caption{(Continued) Comparing the variation of the average cut for different partitioning numbers. The weight of vertices are unit and the weight of hyperedges are their sizes.}
\end{figure}


\begin{figure}[t]
	\centering
    \ContinuedFloat 

	\begin{subfigure}{0.45\textwidth}
		\centering
		\captionsetup{font=small,labelfont=bf}
		\resizebox {\textwidth} {!} {
		\fbox{\includegraphics{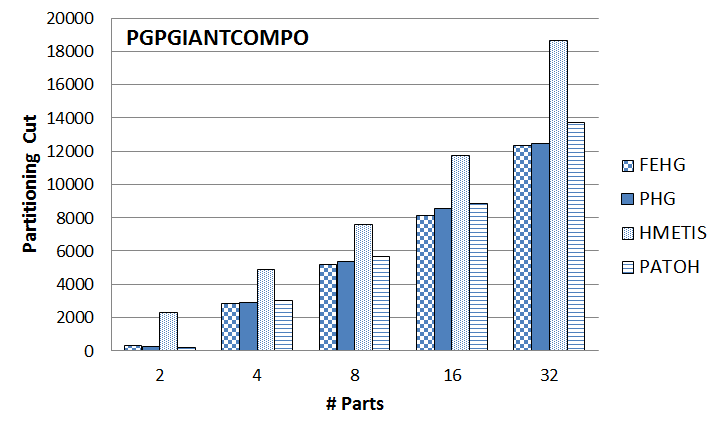}}
		}
	\caption{}
	\end{subfigure}
	
	\begin{subfigure}{0.45\textwidth}
		\centering
		\captionsetup{font=small,labelfont=bf}
		\resizebox {\textwidth} {!} {
		\fbox{\includegraphics{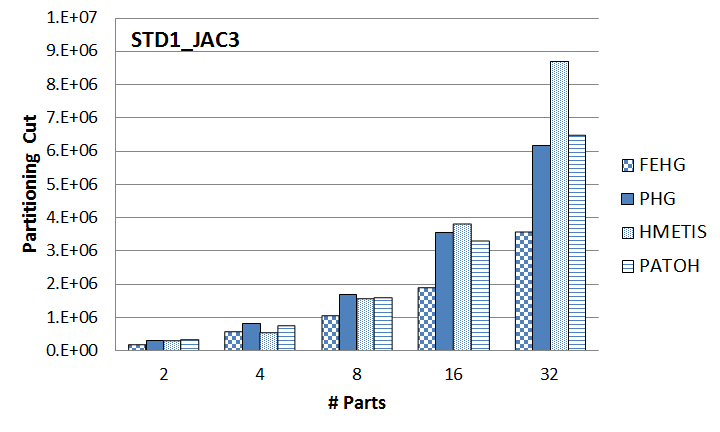}}
	}
	\caption{}
	\end{subfigure}

\caption{(Continued) Comparing the variation of the average cut for different partitioning numbers. The weight of vertices are unit and the weight of hyperedges are their sizes.}
\end{figure}

In some practical applications such as parallel distributed systems, hypergraph partitioning is employed to reduce the communication volume between the processor set. In this situation, the weight of hyperedges represents the volume of communications between a group of vertices, and the objective of  hypergraph partitioning is to reduce the number of messages communicated between processors as well as the volume of communications\footnote{If we assume the vertices are the tasks of the parallel application, the weight of the vertices shows the amount of computational effort the processors spend for each vertex. In our scenario, we assumed that the computation time spent for processing all vertices is the same (unit vertex weights) and the aim is to reduce the number and volume of communications. An example of this situation is in large scale vertex-centric graph processing tools such as Pregel \cite{malewicz2010}.}. In order to model this scenario, we set the weight of the hyperedges to be their sizes for the next phase of simulations. The main reason for this simulation is that we want to investigate the performance of the clustering algorithms in multi-level hypergraph partitioning tools when there are weights on the hyperedges. 

When hyperedges have different weights, vertex connectivity is no longer the only measure used for clustering decisions. Compared to the previous scenario, taking a group of strongly connected components of vertices will not always result in cut reduction as connectivity, as well as how tightly the vertices are connected to each other, is important. The simulation results for this scenario are depicted in Fig.~\ref{fig:partcut_scale_edge}. 

According to the results, \textit{FEHG} gives the best partitioning cut on most of the hypergraphs (in 40 out of the 55 cases). In our evaluation, we have three different types of hypergraphs. The first group are those with very irregular structure and high variation of vertex degree or hyperedge size:  \texttt{CNR-2000}, \texttt{GUPTA1}, \texttt{Notredame\_WWW}, \texttt{AS-22JULY06}, and \texttt{STD\_JAC3}. \textit{FEHG} gives much better quality, finding smaller cuts than the other algorithms in every case. This shows that \textit{FEHG} suits well this type of hypergraph (such as is found in social networks). 

The second group have less irregularity:  \texttt{COND-MAT-2005}, \texttt{PGPGIANTCOMPO}, and \texttt{CELEGANSNEURAL} hypergraphs. These hypergraphs have less variable vertex degree or hyperedge size than the first group. Again, \textit{FEHG} gives the best partitioning results on these types of hypergraphs, but the difference between all partitioners, except \textit{hMetis}, is small. On these types of hypergraphs, we can get reasonable partitioning quality using local partitioners and the performance of the algorithm highly depends on the vertex similarity measure; for example, the one proposed by \textit{hMetis} gives the worst quality. 

The third group are those with regular structure and very small variability of vertex degree and hyperedge size: \texttt{NETSCIENCE}, \texttt{PATENTS\_MAIN}, and \texttt{MARK3JAC120}. The evaluations show that the quality of \textit{FEHG} is worse than the other partitioners. In the case of \texttt{NETSCIENCE} (which has a very small size), most of the algorithms go through only one level of coarsening. The difference between the cuts is less than 50. Due to the regular structure, local vertex matching decisions give much better results than global vertex matching. We have noticed that in these hypergraphs, the algorithm builds cores that contain a very small fraction of the vertices. Therefore, \textit{FEHG} mostly relies on the local and random vertex matching which is based on \textit{Jaccard} similarity. It seems that \textit{Jaccard} similarity does not perform well compared to the other partitioners and the agglomerative vertex matching of \textit{PHG} gives the best results. 

The results show that \textit{PaToH}, which was very competitive with our algorithm for the unit hyperedge size tests, here generates very bad partitioning results. This suggests that our algorithm is more reliable than \textit{PaToH} considered over all types of hypergraphs. Overall, \textit{PaToH} and, then, \textit{hMetis} generate the worst partitioning quality. Some of the partitioning results are not reported for \textit{hMetis} because the algorithm  terminates with an internal error on some of the hypergraphs and part numbers. Perhaps the reason is that \textit{hMetis} suits only partitioning on unit hyperedge size as it is designed for VLSI circuit partitioning.

The running times of the algorithms are reported in Table~\ref{tab:weighed_runtime}. The ranking of algorithms in order of decreasing running time is \textit{hMetis}, \textit{FEHG}, \textit{PHG}, and \textit{PaToH}. 
We note that allowing multiple matches of the vertices can provide not only better partitioning quality compared to pair-matching, but also that it can improve the running time of the algorithm because of the faster reduction in hypergraph size during coarsening. In the case of non-unit hyperedge weights, we have tested our algorithm to see the effects of multiple matching on the performance of the algorithm. In order to do this, we match all vertices that belong to a core when a core is found in our rough set clustering algorithm. The only limitation is that we do not allow the weight of a coarser vertex to exceed the size of a part because it makes it difficult to maintain the balance constraint. The evaluation shows that using multiple matching in our algorithm can improve the runtime by up to 7\% and the maximum improvement is observed for \texttt{CNR-2000} (up to 30\% improvement in runtime). 

\setlength{\extrarowheight}{1pt}

\begin{table*}[t]
	\captionsetup{font=small,labelfont=bf}
	\caption{Comparing the running time of algorithms for different partitioning. Vertices have unit weights and hyperedge weights are equal to their size. The times are reported in milliseconds.}\label{tab:weighed_runtime}
	\centering
	\resizebox{0.7\textwidth}{!}{%
	\begin{threeparttable}
		\begin{tabular}{|l|c|ccccc|}
			\hline
			& &\multicolumn{5}{|c|}{\textbf{Number of Parts}} \\ 
			\cline{3-7}
			& & \textbf{2} & \textbf{4} & \textbf{8} & \textbf{16} & \textbf{32}\\
			\hline
			
			& FEHG-ADJ & 109  & 210	& 308	& 412	& 523	\\
			\textbf{ AS-22JULY06}
			& PHG 		 & 157	& 274	& 413	& 522	& 634\\
			& hMetis	 & 126	& 344	& 803	& 1370	& 5902	\\
			& PaToH	 & 82	& 212	& 336	& 422	& 514	\\
			\hline

			 &FEHG-ADJ& 	8& 	15& 	21& 27& 	33\\
			\textbf{ CELEGANSNEURAL}
 	  		 &PHG 	 &	4& 	7& 	19& 	25& 	22\\
			 &HMETIS	 & 	12& 	18& 	32& 	--& 	--\\
			 &PATOH& 	4& 	4& 	6& 	8& 	12\\
			\hline

			&FEHG-ADJ& 	5& 	10& 	17& 	27& 34\\
			\textbf{ NETSCIENCE}
			&PHG	& 	4& 	6& 	10& 	22& 	32\\
			&HMETIS	& 	--& 	--& 	14& 	20& 	--\\
			&PATOH	& 	2& 	2& 	4& 	4& 	8\\
			\hline

			&FEHG-ADJ& 	114& 	224& 	325& 	408& 	491\\
			\textbf{ PGPGIANTCOMPO}
			&PHG	& 	44& 	57& 	89& 	114& 	147\\
			&HMETIS	& 	170& 	234& 	354& 	452& 	544\\
			&PATOH	& 	12& 	20& 	32& 	46& 	62\\
			\hline

			&FEHG-ADJ& 	19480& 	30570& 	39720& 	50140& 	57560\\
			\textbf{ CNR-2000}
			&PHG	& 	3035& 	5202& 	7317& 	9267& 	11060\\
			&HMETIS	& 	22590& 	41680& 	50990& 	61190& 	68850\\
			&PATOH	& 	2004& 	3960& 	6000& 	8084& 	10390\\
			\hline

			&FEHG-ADJ& 	1843& 	3014& 	4020& 	4918& 	6095\\
			\textbf{ GUPTA1}
			&PHG	& 	937& 	1853& 	2648& 	3453& 	4285\\
			&HMETIS	& 	994& 	4066& 	11990& 	43000& 	331000\\
			&PATOH	& 	914& 	2140& 	3544& 	5370& 	7298\\
			\hline

			&FEHG-ADJ& 	708& 	1304& 	1913& 	2546& 	3192\\
			\textbf{ MARK3JAC120}
			&PHG	& 	318& 	588& 	891& 	1204& 	1592\\
			&HMETIS	& 	1748& 	4570& 	7010& 	9410& 	11130\\
			&PATOH	& 	128& 	272& 	416& 	604& 	796\\
			\hline

			&FEHG-ADJ& 	1588& 	4071& 	6487& 	9095& 	11130\\
			\textbf{ NOTREDAME\_WWW}
			&PHG	& 	2129& 	3673& 	5054& 	6203& 	7207\\
			&HMETIS	& 	5442& 	12770& 	17190& 	23270& 	28060\\
			&PATOH	& 	632& 	1262& 	1950& 	2626& 	3316\\
			\hline

			&FEHG-ADJ& 	1933& 	3187& 	4430& 	5860& 	7514\\
			\textbf{ PATENTS\_MAIN}
			&PHG	& 	1274& 	2156& 	2919& 	3610& 	4251\\
			&HMETIS	& 	11850& 	24080& 	32860& 	38580& 	42630\\
			&PATOH	& 	396& 	734& 	1024& 	1340& 	1648\\
			\hline

			&FEHG-ADJ& 	4970& 	12270& 	19610& 	26710& 	32630\\
			\textbf{ STD1\_JAC3}
			&PHG	 & 	1116& 	2005& 	2775& 	3451& 	4033\\
			&HMETIS	 & 	4086& 	11480& 	19610& 	57300& 	175500\\
			&PATOH	 & 	1720& 	3884& 	5372& 	8380& 	10830\\
			\hline

			&FEHG-ADJ& 	643& 	1137& 	1612& 	2210& 	2772\\
			\textbf{ COND-MAT-2005}
			&PHG	& 	318& 	535& 	750& 	954& 	1178\\
			&HMETIS	& 	3800& 	7038& 	9930& 	13740& 	20020\\
			&PATOH	& 	162& 	284& 	370& 	500& 	584\\

			\hline
		\end{tabular}
	\end{threeparttable}
	}	
\end{table*}

In another evaluation, we have evaluated the performance of our clustering coefficient update strategy. For this purpose, we calculate the CC of the hypergraph in every coarsening level and compare the results with when updates are used. The quality does not improve in all cases. For example, the quality of the third type of hypergraph described above was diminished by 1\% on average. The best quality improvement is for \texttt{CNR-2000} that is 6\% and it was between 0.2\% to 1.5\% on other hypergraphs. On the other hand, the runtime of the algorithms are increased by up to 16\%. This shows that our update method is very reliable and there is no need to calculate the CC in each coarsening level. 

\begin{table*}[t]
	\captionsetup{font=small,labelfont=bf}
	\caption{The time that \textit{FEHG} algorithm spends in each phase of the algorithms. Times are reported in seconds.}\label{tab:timing_details}
	\centering
	\resizebox{0.7\textwidth}{!}{%
	\begin{threeparttable}
		\begin{tabular}{|l|c|cccccccc|}
			\hline
			\rot[90]{\textbf{Parts}} &
			& \rot[80]{\textbf{\small AS-22JULY06}}
			& \rot[80]{\textbf{\small CELEGANSNEURAL}}
			& \rot[80]{\textbf{\small NETSCIENCE}}
			& \rot[80]{\textbf{\small PGPGIANTCOMPO}}
			& \rot[80]{\textbf{\small NOTREDAME}}
			& \rot[80]{\textbf{\small PATENTS\_MAIN}}
			& \rot[80]{\textbf{\small STD1\_JAC3}}
			& \rot[80]{\textbf{\small COND-MAT-2005}}\\
			\hline
			
			& Overall	& 0.1461 & 0.0080 & 0.0059 & 0.0831 & 1.5562 & 1.9312 & 7.3650 & 0.6155 \\
			& Build		& 0.0230 & 0.0003 & 0.0014 & 0.0118 & 0.4568 & 0.3496 & 0.4181 & 0.0697\\
			& Recursion	& 0.0000 & 0.0000 & 0.0000 & 0.0000 & 0.0000 & 0.0000 & 0.0000 & 0.0000\\
\textbf{2}
			& Vcycle		& 0.0036 & 0.0000 & 0.0003 & 0.0025 & 0.0048 & 0.0604 & 1.7775 & 0.0203\\
			& HCG	& 0.0224 & 0.0034 & 0.0007 & 0.0257 & 0.0000 & 0.4512 & 3.6833 & 0.2475\\
			& Matching	& 0.0352 & 0.0000 & 0.0009 & 0.0071 & 0.0000 & 0.2275 & 0.1238 & 0.0638\\
			& Coarsening	& 0.0332 & 0.0000 & 0.0013 & 0.0181 & 0.0000 & 0.4377 & 1.1945 & 0.1108\\
			& InitPart	& 0.0190 & 0.0040 & 0.0003 & 0.0124 & 1.0467 & 0.3019 & 0.0466 & 0.0748\\
			& Refinement	& 0.0086 & 0.0000 & 0.0007 & 0.0051 & 0.0286 & 0.0868 & 0.1072 & 0.0260\\
			\hline

			& Overall	& 0.3722 & 0.0218 & 0.0167 & 0.2456 & 6.2414 & 4.9619 & 18.2084 & 1.7036\\
			& Build		& 0.0235 & 0.0009 & 0.0006 & 0.0113 & 0.5750 & 0.3474 & 0.4182 & 0.0700\\
			& Recursion	& 0.0161 & 0.0008 & 0.0026 & 0.0081 & 0.2712 & 0.2091 & 0.4072 & 0.0562\\
\textbf{8}
			& Vcycle		& 0.0095 & 0.0009 & 0.0004 & 0.0088 & 0.1453 & 0.1625 & 4.8115 & 0.0604\\
			& HCG	& 0.0514 & 0.0023 & 0.0049 & 0.0727 & 1.7675 & 1.1641 & 8.7172 & 0.6657\\
			& Matching	& 0.0903 & 0.0005 & 0.0012 & 0.0184 & 0.7289 & 0.5581 & 0.3300 & 0.1749\\
			& Coarsening	& 0.0838 & 0.0013 & 0.0043 & 0.0470 & 1.3447 & 1.0932 & 3.0456 & 0.3349\\
			& InitPart	& 0.0650 & 0.0116 & 0.0002 & 0.0553 & 1.1754 & 1.1883 & 0.1084 & 0.2346\\
			& Refinement	& 0.0309 & 0.0033 & 0.0024 & 0.0230 & 0.2041 & 0.2150 & 0.3463 & 0.1017\\
			\hline

			& Overall	& 0.6258 & 0.0360 & 0.0363 & 0.4007 & 9.8867 & 7.6629 & 28.5887 & 2.7925\\
			& Build		& 0.0233 & 0.0009 & 0.0010 & 0.0110 & 0.4578 & 0.3456 & 0.4173 & 0.0699\\
			& Recursion	& 0.0331 & 0.002 & 0.0027 & 0.0157 & 0.5302 & 0.3908 & 0.8082 & 0.112\\
\textbf{32}
			& Vcycle		& 0.0156 & 0.0013 & 0.0024 & 0.0159 & 0.2789 & 0.2547 & 9.6070 & 0.0992\\
			& HCG	& 0.0776 & 0.0023 & 0.0029 & 0.1059 & 2.9710 & 1.8239 & 11.8124 & 0.9619\\
			& Matching	& 0.1317 & 0.0004 & 0.0028 & 0.0292 & 1.1691 & 0.8517 & 0.4540 & 0.2597\\
			& Coarsening	& 0.1267 & 0.0012 & 0.0054 & 0.0771 & 2.5044 & 1.7197 & 4.3973 & 0.5536\\
			& InitPart	& 0.1372 & 0.0230 & 0.0133 & 0.0905 & 1.4581 & 1.8681 & 0.3006 & 0.4734\\
			& Refinement	& 0.0772 & 0.0045 & 0.0053 & 0.0535 & 0.4734 & 0.3724 & 0.7559 & 0.2545\\
			\hline

		\end{tabular}
	\end{threeparttable}
	}	
\end{table*}

Finally, the detailed running time of \textit{FEHG} and the amount of time the algorithm spends in each section is given in Table~\ref{tab:timing_details} for $\left\lbrace  2,8,32 \right\rbrace$-way partitioning on some of the hypergraphs. In the table, the overall running time is given in the first row. \textit{Build} is the time for building data structures and preparation time, \textit{recursion} is recursive bipartitioning time, \textit{vcycle} is the amount of time for reduction and hypergraph projection in the multi-level paradigm, \textit{HCG} is for building HCG, \textit{matching} includes the time for rough set matching algorithm. Finally, 
\textit{coarsening, initPart} and \textit{refinement} are the time taken for building the coarser hypergraph in the coarsening phase, initial partitioning and uncoarsening phases of \textit{FEHG}. 

The most time consuming part of the algorithm is building HCG: around $27\%$ of the whole running time. The rough set clustering takes only $13\%$ of the runtime. Building the coarser hypergraph and the initial partitioning and the coarsening each takes around $20\%$. One can reduce the initial partitioning time by decreasing the number of algorithms in this section. According to the data, the part where one can most usefully perform optimisations is in building the HCG.  If the number of hyperedges is much higher than the number of vertices, its running time can take up most of the algorithm's running time. The refinement phase takes at most $6\%$ of the whole running time. Therefore, using mores passes of the FM algorithm to improve the quality will not increase the overall time significantly. On the other hand, there is little need for this. As discussed in \cite{karytech2002}, a good coarsening algorithm causes less effort in the refinement phase and increasing the passes of the FM algorithm does not make considerable improvement to the cut. This is the case for our \textit{FEHG} algorithm.

\section{Conclusion}\label{sec:conclusion}

In this paper, we have proposed a serial multi-level hypergraph partitioning algorithm (\textit{FEHG}) based on feature extraction and attribute reduction using rough set clustering. The hypergraph is transformed into an information system and rough set clustering techniques are used to find pair-matches of the vertices during the coarsening phase. This was done by, first, categorising the vertices into core and non-core vertices which is a global clustering decision using indispensability relations. In the later step, cores are traversed one at a time to find best matchings between vertices. This provides a trade-off between global and local vertex matching decisions.

The algorithm is evaluated against the state-of-the-art partitioning algorithms and we have shown that \textit{FEHG} can achieve up to 40\% quality improvement on hypergraphs from real applications. We chose our test hypergraphs to model different scenarios in real applications and we found that \textit{FEHG} is the most reliable algorithm and generates the highest quality partitionings on most of the hypergraphs. The quality improvement was much better in hypergraphs with more irregular structures; that is, with higher variation of vertex degree or hyperedge size.

We found that one of the drawbacks of local vertex matching decisions is that they perform very differently under various problem circumstances and their behaviour can change based on the structure of the hypergraph under investigation. The worst case observed was \textit{PaToH} that generated very good and competitive partitioning compared to our algorithm when the hyperedge weights were assumed to be 1, while it gave much worse quality when the hyperedge weights were driven by the hyperedge sizes.

Evaluation of the runtime of the algorithms has shown that the \textit{FEHG}, while using global clustering decisions, runs slower than \textit{PHG} and \textit{PaToH}, but faster than \textit{hMetis}. We showed that the runtime can be improved by using multiple matching on the vertex set instead of pair-matching. Furthermore, we have observed that the most time consuming part of the algorithm is building HCG. In future work, we are planning to improve this aspect of the proposed algorithm to improve the runtime. 

The performance of serial hypergraph partitioning algorithms is limited and it is not possible to partition very large hypergraphs with  billions of vertices and hyperedges using the computing resources of one computer, and in future work we will propose a scalable parallel version of the \textit{FEHG} algorithm based on the parallel rough set clustering techniques.

\begin{acknowledgements}
We would like to thank four anonymous reviewers who gave helpful comments on a preliminary version of this paper. We also thank Prof. Andre Brinkmann and Dr. Lars Nagel from the Efficient Computing and Storage Group at Johannes Gutenberg University of Mainz for their help and support.
\end{acknowledgements}

\bibliographystyle{spmpsci}      
\bibliography{references}{}

\begin{thebibliography}{10}
\providecommand{\url}[1]{{#1}}
\providecommand{\urlprefix}{URL }
\expandafter\ifx\csname urlstyle\endcsname\relax
  \providecommand{\doi}[1]{DOI~\discretionary{}{}{}#1}\else
  \providecommand{\doi}{DOI~\discretionary{}{}{}\begingroup
  \urlstyle{rm}\Url}\fi

\bibitem{alpert2000fixed}
Alpert, C., Caldwell, A., Kahng, A., Markov, I.: Hypergraph partitioning with
  fixed vertices [vlsi cad].
\newblock Computer-Aided Design of Integrated Circuits and Systems, IEEE
  Transactions on \textbf{19}(2), 267--272 (2000)

\bibitem{alp1996}
Alpert, C.J.: Multi-way graph and hypergraph partitioning.
\newblock Ph.D. thesis, UCLA Computer Science Department (1996)

\bibitem{alpert1998multilevel}
Alpert, C.J., Huang, J.H., Kahng, A.B.: Multilevel circuit partitioning.
\newblock Computer-Aided Design of Integrated Circuits and Systems, IEEE
  Transactions on \textbf{17}(8), 655--667 (1998)

\bibitem{aykanat2008fixed}
Aykanat, C., Cambazoglu, B.B., U{\c{c}}ar, B.: Multi-level direct k-way
  hypergraph partitioning with multiple constraints and fixed vertices.
\newblock Journal of Parallel and Distributed Computing \textbf{68}(5),
  609--625 (2008)

\bibitem{bloznelis2013}
Bloznelis, M., et~al.: Degree and clustering coefficient in sparse random
  intersection graphs.
\newblock The Annals of Applied Probability \textbf{23}(3), 1254--1289 (2013)

\bibitem{ccatalyurek2011patoh}
{\c{C}}ataly{\"u}rek, {\"U}., Aykanat, C.: Patoh (partitioning tool for
  hypergraphs).
\newblock In: Encyclopedia of Parallel Computing, pp. 1479--1487. Springer
  (2011)

\bibitem{catayk1999}
Catalyurek, U.V., Aykanat, C.: Hypergraph-partitioning-based decomposition for
  parallel sparse-matrix vector multiplication.
\newblock Parallel and Distributed Systems, IEEE Transactions on
  \textbf{10}(7), 673--693 (1999)

\bibitem{cong1998}
Cong, J., Lim, S.K.: Multiway partitioning with pairwise movement.
\newblock In: Proceedings of the 1998 IEEE/ACM International Conference on
  Computer-aided Design, ICCAD '98, pp. 512--516. ACM, New York, NY, USA (1998)

\bibitem{curino2010schism}
Curino, C., Jones, E., Zhang, Y., Madden, S.: Schism: a workload-driven
  approach to database replication and partitioning.
\newblock Proceedings of the VLDB Endowment \textbf{3}(1-2), 48--57 (2010)

\bibitem{sparsecollection2011}
Davis, T.A., Hu, Y.: The university of florida sparse matrix collection.
\newblock ACM Transactions on Mathematical Software \textbf{38}(1), 1 (2011)

\bibitem{devetal2006}
Devine, K., Boman, E., Heaphy, R., Bisseling, R., Catalyurek, U.: Parallel
  hypergraph partitioning for scientific computing.
\newblock In: Parallel and Distributed Processing Symposium, 2006. IPDPS 2006.
  20th International, p. 10pp (2006)

\bibitem{erdos1966representation}
Erdos, P., Goodman, A.W., P{\'o}sa, L.: The representation of a graph by set
  intersections.
\newblock Canad. J. Math \textbf{18}(106-112), 86 (1966)

\bibitem{ertoz2002new}
Ertoz, L., Steinbach, M., Kumar, V.: A new shared nearest neighbor clustering
  algorithm and its applications.
\newblock In: Workshop on Clustering High Dimensional Data and its Applications
  at 2nd SIAM International Conference on Data Mining, pp. 105--115 (2002)

\bibitem{ertozetal2003}
Ert{\"o}z, L., Steinbach, M., Kumar, V.: Finding clusters of different sizes,
  shapes, and densities in noisy, high dimensional data.
\newblock In: SDM, pp. 47--58. SIAM (2003)

\bibitem{fm1982}
Fiduccia, C.M., Mattheyses, R.M.: A linear-time heuristic for improving network
  partitions.
\newblock In: 19th Conference on Design Automation, pp. 175--181. IEEE (1982)

\bibitem{fjall1998graphsurvey}
Fj{\"a}llstr{\"o}m, P.O.: Algorithms for graph partitioning: A survey.
\newblock Link{\"o}ping electronic articles in computer and information science
  \textbf{3}(10) (1998)

\bibitem{foudalis2011social}
Foudalis, I., Jain, K., Papadimitriou, C., Sideri, M.: Modeling social networks
  through user background and behavior.
\newblock In: Proceedings of the 8th International Conference on Algorithms and
  Models for the Web Graph, WAW'11, pp. 85--102. Springer-Verlag (2011)

\bibitem{garey1979computers}
Garey, M.R., Johnson, D.S.: Computers and intractability: a guide to the theory
  of np-completeness. 1979.
\newblock San Francisco, LA: Freeman  (1979)

\bibitem{gavin2002functional}
Gavin, A.C., B{\"o}sche, M., Krause, R., Grandi, P., Marzioch, M., Bauer, A.,
  Schultz, J., Rick, J.M., Michon, A.M., Cruciat, C.M., et~al.: Functional
  organization of the yeast proteome by systematic analysis of protein
  complexes.
\newblock Nature \textbf{415}(6868), 141--147 (2002)

\bibitem{karytech2002}
George, K.: Multilevel hypergraph partitioning.
\newblock Tech. rep., University of Minnesota (2002)

\bibitem{goldberg1983}
Goldberg, M.K., Burstein, M.: Heuristic improvement technique for bisection of
  VLSI networks.
\newblock IBM Thomas J. Watson Research Division (1983)

\bibitem{bey2014}
Heintz, B., Chandra, A.: Beyond graphs: Toward scalable hypergraph analysis
  systems.
\newblock SIGMETRICS Perform. Eval. Rev. \textbf{41}(4), 94--97 (2014)

\bibitem{hendrickson1998graph}
Hendrickson, B.: Graph partitioning and parallel solvers: Has the emperor no
  clothes?
\newblock In: Solving Irregularly Structured Problems in Parallel, pp.
  218--225. Springer (1998)

\bibitem{hu2014}
Hu, T., Liu, C., Tang, Y., Sun, J., Xiong, H., Sung, S.Y.: High-dimensional
  clustering: a clique-based hypergraph partitioning framework.
\newblock Knowledge and information systems \textbf{39}(1), 61--88 (2014)

\bibitem{ihler1993}
Ihler, E., Wagner, D., Wagner, F.: Modeling hypergraphs by graphs with the same
  mincut properties.
\newblock Inf. Process. Lett. \textbf{45}(4), 171--175 (1993)

\bibitem{hmetis15url}
Karypis, G.: hmetis: Hypergraph and circuit partitioning - version 1.5 (2007)

\bibitem{karypis1998hmetis}
Karypis, G., Kumar, V.: hmetis: A hypergraph partitioning package version 1.5",
  user manual (1998)

\bibitem{karypis1999kway}
Karypis, G., Kumar, V.: Multilevel k-way hypergraph partitioning.
\newblock In: Proceedings of ACM/IEEE Design Automation Conference, pp.
  343--348 (1999)

\bibitem{klamt2009hypergraphs}
Klamt, S., Haus, U.U., Theis, F.: Hypergraphs and cellular networks.
\newblock PLoS Comput Biol \textbf{5}(5), e1000,385 (2009)

\bibitem{latapy2008basic}
Latapy, M., Magnien, C., Del~Vecchio, N.: Basic notions for the analysis of
  large two-mode networks.
\newblock Social Networks \textbf{30}(1), 31--48 (2008)

\bibitem{malewicz2010}
Malewicz, G., Austern, M.H., Bik, A.J., Dehnert, J.C., Horn, I., Leiser, N.,
  Czajkowski, G.: Pregel: A system for large-scale graph processing.
\newblock In: Proceedings of the 2010 ACM SIGMOD International Conference on
  Management of Data, SIGMOD '10, pp. 135--146. ACM, New York, NY, USA (2010)

\bibitem{marquez2015}
Márquez, C., Cesar, E., Sorribes, J.: Graph-based automatic dynamic load
  balancing for hpc agent-based simulations.
\newblock In Proc. of 3rd Workshop on Parallel and Distributed Agent-Based
  Simulations (PADABS2015), Vienna, Austria  (2015)

\bibitem{pawlak1991}
Pawlak, Z.: Rough Sets: Theoretical Aspects of Reasoning About Data.
\newblock Kluwer Academic Publishers, Norwell, MA, USA (1991)

\bibitem{pawlak2005rough}
Pawlak, Z., Polkowski, L., Skowron, A.: Rough sets: An approach to vagueness.
\newblock Encyclopedia of Database Technologies and Applications pp. 575--580
  (2005)

\bibitem{saab1992rao}
Saab, Y., Rao, V.: On the graph bisection problem.
\newblock Circuits and Systems I: Fundamental Theory and Applications, IEEE
  Transactions on \textbf{39}(9), 760--762 (1992)

\bibitem{san1989kfm}
Sanchis, L.: Multiple-way network partitioning.
\newblock Computers, IEEE Transactions on \textbf{38}(1), 62--81 (1989)

\bibitem{trilinosurl}
{Sandia National Laboratories}: Trilinos: Open source software libraries for
  the development of scientific applications (2014)

\bibitem{zoltanurl}
{Sandia National Laboratories}: Zoltan: Parallel partitioning, load balancing
  and data-management services (2014)

\bibitem{skowron1992}
Skowron, A., Rauszer, C.: The discernibility matrices and functions in
  information systems.
\newblock In: Intelligent Decision Support, pp. 331--362. Springer (1992)

\bibitem{steinbach2000}
Steinbach, M., Karypis, G., Kumar, V., et~al.: A comparison of document
  clustering techniques.
\newblock KDD workshop on text mining \textbf{400}(1), 525--526 (2000)

\bibitem{roughreview2009}
Thangavel, K., Pethalakshmi, A.: Review: Dimensionality reduction based on
  rough set theory: A review.
\newblock Appl. Soft Comput. \textbf{9}(1), 1--12 (2009)

\bibitem{tian2009}
Tian, Z., Hwang, T., Kuang, R.: A hypergraph-based learning algorithm for
  classifying gene expression and arraycgh data with prior knowledge.
\newblock Bioinformatics \textbf{25}(21), 2831--2838 (2009)

\bibitem{ale2006}
Trifunovic, A.: Parallel algorithms for hypergraph partitioning.
\newblock Ph.D. thesis, University of London (2006)

\bibitem{wang2014bilionnode}
Wang, L., Xiao, Y., Shao, B., Wang, H.: How to partition a billion-node graph.
\newblock In: Data Engineering (ICDE), 2014 IEEE 30th International Conference
  on, pp. 568--579. IEEE (2014)

\bibitem{wroblewski1995}
Wr{\'o}blewski, J.: Finding minimal reducts using genetic algorithms.
\newblock In: Proccedings of the second annual join conference on infromation
  science, pp. 186--189 (1995)

\bibitem{wroblewski1998}
Wr{\'o}blewski, J.: Genetic algorithms in decomposition and classification
  problems.
\newblock In: Rough Sets in Knowledge Discovery 2, pp. 471--487. Springer
  (1998)

\bibitem{zhou2006learning}
Zhou, D., Huang, J., Sch{\"o}lkopf, B.: Learning with hypergraphs: Clustering,
  classification, and embedding.
\newblock In: Advances in neural information processing systems, pp. 1601--1608
  (2006)

\bibitem{ziarko1995}
Ziarko, W., Shan, N.: Discovering attribute relationships, dependencies and
  rules by using rough sets.
\newblock In: hicss, p. 293. IEEE (1995)

\end{thebibliography}

\end{document}